# Quantitative Predictive Theories through Integrating Quantum, Statistical, Equilibrium, and Nonequilibrium Thermodynamics

Zi-Kui Liu

Department of Materials Science and Engineering, The Pennsylvania State University, University Park, Pennsylvania 16802, USA, email: zxl15@psu.edu

**Abstract:**

Today's thermodynamics is largely based on the combined law for equilibrium systems and statistical mechanics derived by Gibbs in 1873 and 1901, respectively, while irreversible thermodynamics for nonequilibrium systems resides essentially on the Onsager Theorem as a separate branch of thermodynamics developed in 1930s. Between them, quantum mechanics was invented and was quantitatively solved in terms of density functional theory (DFT) in 1960s. These three scientific domains operate based on different principles and are very much separated from each other. In analogy to the parable of the blind men and the elephant articulated by Perdew, they individually represent different portions of a complex system and thus are incomplete by themselves alone, resulting in the lack of quantitative agreement between their predictions and experimental observations. Over the last two decades, the author's group has developed a multiscale entropy approach (recently termed as zentropy theory) that integrates DFT-based quantum mechanics and Gibbs statistical mechanics and is capable of accurately predicting entropy and free energy of complex systems. Furthermore, in combination with the combined law for nonequilibrium systems developed by Hillert, the author developed the theory of cross phenomena beyond the phenomenological Onsager Theorem. The zentropy theory and theory of cross phenomena jointly provide quantitative predictive theories for systems from electronic to any observable scales as reviewed in the present work.

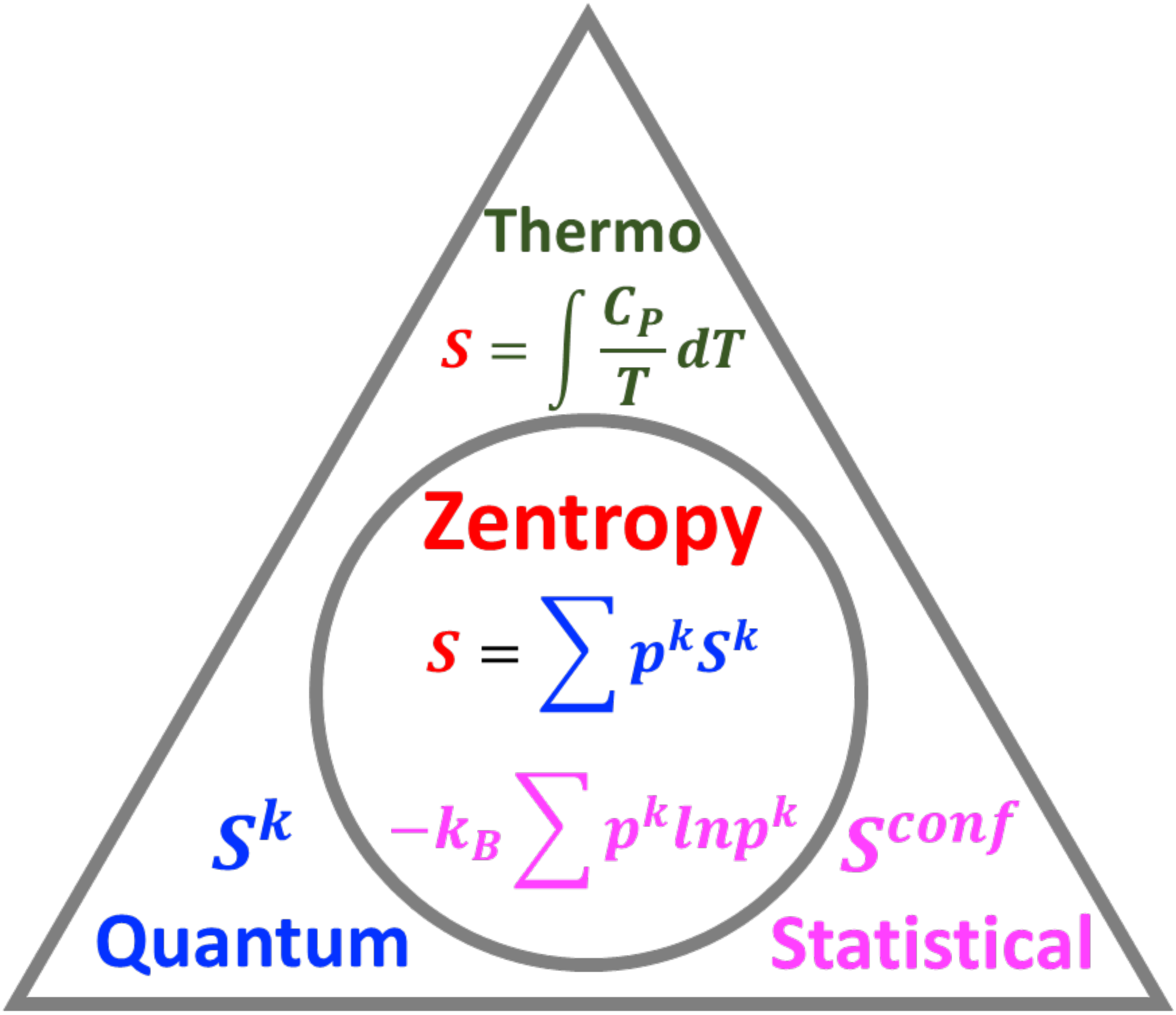

Schematic representation of the zentropy theory for the total entropy $S$ with contributions from each configuration $k$ via DFT calculations ($S^k$) and configurational entropy ($S^{conf} = -k_B \sum p^k lnp^k$) among configurations with the probability $p^k$ for each configuration $k$, which is equal to the entropy from the integration of experimental heat capacity ($C_P$).

# Table of Contents

## 1 Introduction

As written in the preface of "The collected works of J. Willard Gibbs: Vol. I Thermodynamics"[1], Gibbs spent the winter of 1866-67 in Paris and then a year in Berlin to hear the lectures of physics and mathematics.. In 1868 he went to Heidelberg, probably listened to Kirchhoff and Helmholtz, before returning to the US in June 1869. In 1873, Gibbs [2] introduced entropy change into the 1$^{st}$ law of thermodynamics using the relation between entropy change and heat change by Clausius for an reversible process of a closed system under hydrostatic pressure and extended it to multicomponent system in 1875/1876 and 1877/1878 [3,4]. In 1901, in searching for description of absolute entropy, Gibbs [5] took a broader view of statistical mechanics based on the foundation established by Clausius, Maxwell, and Boltzmann and "imagined a great number of systems of the same nature, but differing in the configurations and velocities which they have at a given instant, and differing not merely infinitesimally, but it may be so as to embrace every conceivable combination of configuration and velocities". This definition of configuration is substantially different from the "inquiries … were somewhat narrower in their scope than that which has been mentioned, being applied to the particles of a system, rather than to independent systems." [5] though this narrower approach is still widely used in the literature today. Gibbs thus derived the configurational entropy in terms of the probabilities of the configurations and the statistical mechanics commonly used today. While Gibbs principally focused on thermodynamic and statistical equilibrium in his work, he did mention free energy change for nonequilibrium systems [3,4].

In contrast to the macroscopic top-down view of systems by Gibbs, quantum mechanics developed in 1920s examines the interactions among electrons and atoms and considers a condensed matter as a collection of interactions among nuclei and electrons. Theoretically these interactions can be treated by solving the many-body Schrödinger equation with both nuclei and electrons [6]. The Gibbs classic statistical mechanics was also updated to quantum statistical mechanics [7]. The solutions to the many-body nature of the interactions in quantum mechanics were developed in 1960's in terms of the density functional theory (DFT) that aims to represent the outcome of those interactions and articulates that for a given system, there exists a ground-state configuration at 0 K that its energy is at its minimum value with a universal functional of the interacting electron gas density [8]. The ground-state electron density is obtained by treating the independent-electron kinetic energy and long-range Coulomb interaction energy separately and using independent valence electrons with an exchange-correlation functional of the electron density and an associated exchange-correlation energy to replace the many-body electron interactions [9].

In last 60 years, significant processes have been made in developing the exchange-correlation functionals, particularly the generalized gradient approximation (GGA) [10–12], that treats the exchange-correlation energy as a function of the local electron density and its gradient. The latest meta-GGA includes the strongly constrained and appropriately normed (SCAN) [13,14] and $r^2$SCAN with both improved accuracy and numerical stability and efficiency [14–19]. For property at finite temperature, the entropy of thermal electrons and phonons can be formulated through an interacting inhomogeneous electron gas model [9,20] and the lattice vibrations [21], respectively.

Since the 2nd law of thermodynamics was effectively excluded in Gibbs thermodynamics due to its inequality, this resulted in the development of irreversible thermodynamics phenomenologically [22] based on experimental observations rather than from the 1st and 2nd laws of thermodynamics [23] by Onsager [24,25] in terms of the generalized flux equations and the symmetrical kinetic coefficient matrix and by Prigogine et al.[26–31] in terms of the unified formulation of dynamics and thermodynamics. Nevertheless, Onsager [24] remarked that "the principle of microscopic reversibility is less general than the fundamental laws of thermodynamics", while the present author recently pointed out that it is actually incompatible with irreversible internal processes [32,33]. It is further noted that the microscopic or quantum violation of the 2nd law of thermodynamics has been a topic in the literature forever with some references discussed in Section 5.4 of ref. [32] by the present author, all due to incomplete counting of entropy. A number of recent publications also articulated that quantum correlations do not lead to a violation of the 2nd law of thermodynamics [34,35], and inconsistency presented in the literature is the result of faulty analysis due to flaws in approximations [36,37]. To avoid misunderstanding and confusion, the 2nd law of thermodynamics should be applied to independent internal processes which could be re-defined as systems to study as discussed in Section 3.1 in the present paper.

Over two decades ago, the author's group [21,38–40] started to incorporate predictions from DFT calculations into thermodynamic modeling of multicomponent solutions based on the CALPHAD method where model parameters were mostly evaluated from experimental information [41]. It soon became evident that the DFT-based bottom-up approach using the ground-state configuration alone was not able to quantitively predict free energy of a phase at finite temperature. Consequently, the author's group started to explore quantitative approaches for accurate prediction of the temperature-pressure (T-P) phase diagram including its critical point without fitting parameters through integration of bottom-up DFT and top-down statistical mechanics, using the magnetic transitions in Ce as the first example with two [42] and three [43] configurations, respectively. These configurations include the ground-state and non-ground-state or excited-state configurations and are in accordance with the definition of configurations that Gibbs introduced in his statistical mechanics. By considering the entropy contributions of each configuration predicted by DFT calculations in terms of thermal electrons and phonons and the configurational entropy among the all configurations to the total entropy, the T-P and T-V phase diagrams of Ce were accurately predicted without fitting parameters. This multi-scale entropy approach [44] has been successfully used to accurately predict magnetic phase transition temperatures and property anomalies in a number of magnetic materials [45], including the negative thermal expansion in $Fe_3Pt$ [46] and strongly correlated physics in $YNiO_3$ [47,48]. It was recently termed as the zentropy theory [49].

With the accurate prediction of free energy using the zentropy theory and inspired by the collaboration on thermodiffusion with Murch and his team [50–54], the author [32] revisited the irreversible thermodynamics from the combined 1$^{st}$ and 2$^{nd}$ law of thermodynamics with internal processes emphasized by Hillert [22]. By examining the fundamental flaws of the Onsager theorem and considering the correlations between flux equations as internal processes in a system, the author presented a theory of cross phenomena based on combined 1$^{st}$ and 2$^{nd}$ law of thermodynamics that captures the experimental observations that the Onsager theorem aimed to phenomenologically represent and enables the prediction of cross phenomena coefficients as 2nd derivatives of free energy [32,33].

In this overview paper, the zentropy theory and theory of cross phenomena will be presented along with potential future improvements and applications. The rest of the paper is organized as follows: Section 2 for overview of Gibbs equilibrium thermodynamics, DFT-based quantum mechanics, and statistical mechanics; Section 3 for Hillert nonequilibrium thermodynamics, Onsager theorem, Prigogine entropy balance, and Agren atomic mobility; Section 4 for the zentropy theory; Section 5 for the theory of cross phenomena; and Section 6 for summary and future outlooks.

# 2 Gibbs equilibrium thermodynamics, DFT-based quantum mechanics, and statistical mechanics

## 2.1 Gibbs equilibrium thermodynamics

Employing the relation between entropy change and heat change by Clausius for an reversible process of a closed system under hydrostatic pressure, Gibbs [2] replaced the heat change in the 1st law of thermodynamics by the entropy change and obtained the combined law of thermodynamics for a closed *equilibrium* system, which is written in today's convention as follows

$$dU = dQ + dW = TdS - PdV \qquad Eq.\ 1$$

where $dU$ is the internal energy change of the system, $dQ$, $dW$, $dS$ and $dV$ are the increases of heat, work, entropy, and volume of the system controlled from the surroundings, and $T$ and $P$ are the temperature and pressure. Even though Eq. 1 is commonly referred to as the combined law of thermodynamics, it does not contain the key concept of the 2nd law of thermodynamics, i.e., the positive entropy production from irreversible internal processes ($ip$) represented by the following equation

$$d_{ip}S > 0 \qquad \textit{Eq. 2}$$

It is thus evident that Eq. 1 is valid only for $d_{ip}S \leq 0$, i.e., equilibrium systems without any possible internal processes by itself.

In a subsequent paper in the same year, Gibbs [55] emphasized that the relation between U, S, and V carries more knowledge than the other combinations of quantities in Eq. 1 such as V, P and T proposed by Thomson in 1871. Gibbs [56] termed Eq. 1 as the fundamental equation, representing the internal energy as a function of S and V, i.e., U(S,V), with S and V as the natural variables of U [22]. Gibbs [55] further articulated that at equilibrium T and P are both homogeneous in all phases of solid, liquid, and vapor. Gibbs' publications prompted Maxwell to make two three-dimensional models to represent the internal energy surface as a function of entropy and volume with one copy sent to Gibbs [1] and one kept in Cavendish Laboratory at the University of Cambridge [33].

In his next long and more well-known publication in 1878, Gibbs [56] discussed the conditions of equilibrium for heterogeneous masses and introduced the concept of chemical potential with Eq. 1 revised as follows for a system with c independent components in today's convention

$$dU = TdS - PdV + \sum_{i=1}^{c} \mu_i dN_i = \sum_{a=1}^{c+2} Y^a dX^a \qquad \textit{Eq. 3}$$

where $\mu_i$ is the chemical potential of component $i$, $N_i$ the moles of component $i$, and $Y^a$ and $X^a$ represent the pairs of conjugate variables with $Y^a$ for potentials, such as $T$, $P$, and $\mu_i$, and $X^a$ for molar quantities, such as $S$, $V$, and $N_i$ [22,33,57]. As it will be shown later in this paper, $\mu_i$ can be introduced in a way to better define its physical significance. It is noted that Gibbs called $\mu_i$ potentials and considered pressure as a potential too when discussing equilibrium related to surfaces, which are generalized to include temperature plus stress, electric, and magnetic fields in general [22,33,57].

Gibbs [56] elegantly demonstrated that every potential, $Y^a$, is homogeneous in the system with multiple phases in equilibrium with each other and further derived the integral form of Eq. 3 as follows for a homogeneous system

$$U = TS - PV + \sum_{i=1}^{c} \mu_i N_i = \sum_{a=1}^{c+2} Y^a X^a \qquad \textit{Eq. 4}$$

In the same paper, Gibbs [56] introduced the now-termed free energies such as Helmholtz energy ($F = U - TS$), enthalpy ($H = U + PV$), and Gibbs energy ($G = U - TS + PV = \sum_{i=1}^{c} \mu_i N_i$) and re-wrote Eq. 3 accordingly, e.g., Gibbs energy as follows,

$$dG = -SdT + VdP + \sum_{i=1}^{c} \mu_i dN_i \qquad \textit{Eq. 5}$$

Gibbs [56] further presented the now-termed Gibbs-Duhem equation as follows

$$d\Phi = d\left(G - \sum_{i=1}^{c} \mu_i N_i\right) = -SdT + VdP - \sum_{i=1}^{c} N_i d\mu_i = 0 \qquad \textit{Eq. 6}$$

along with the now-termed Gibbs phase rule for an equilibrium with $p$ co-existing phases and the now-termed Clausius-Clapeyron equation in terms of temperature and pressure between two phases in equilibrium

$$\upsilon = c + 2 - p \qquad \textit{Eq. 7}$$

$$\frac{dP}{dT} = \frac{\Delta S}{\Delta V} = \frac{\Delta H}{T\Delta V} \qquad \textit{Eq. 8}$$

where $\upsilon$ represents the number of *independent potentials* for such an equilibrium, and $\Delta S$, $\Delta V$, and $\Delta H$ are the differences of $S$, $V$, and $H$ between the two phases. While Gibbs [56] did not explicitly discuss the entropy production due to internal processes, but showed $dU > 0$ , $dG > 0$, and $d\Phi > 0$ for a stable equilibrium and further wrote

$$\frac{\partial Y^a}{\partial X^a} > 0 \qquad \textit{Eq. 9}$$

where all the other natural variables are kept constant. All partial derivatives in this present paper is written under such a convention unless a clarification is needed.

For the limit of stability approached from the stable region of the system, one has $\frac{\partial Y^a}{\partial X^a} = +0$ and $\frac{\partial X^a}{\partial Y^a} = +\infty$, i.e., the singularity or divergency of properties, and a critical point is reached with $\frac{\partial Y^a}{\partial X^a} = \frac{\partial^2 Y^a}{\partial (X^a)^2} = 0$. As will be shown later in this paper, $d\Phi$ is directly related to entropy production due to internal processes in nonequilibrium systems, and the author suggested to name $\Phi$ as *Duhem energy* [33]. As will be shown below, it may be better to name it as *Hillert energy*. Furthermore, Hillert [22] showed that the partial derivative by Eq. 9 with more potentials as natural variables is smaller than the one with more molar quantities as natural variables. When all other potentials are used as natural variables, one obtains the Gibbs-Duhem equation as shown by Eq. 6.

## 2.2 DFT-based quantum mechanics

DFT represents the state-of-the-art solution of the multi-body Schrödinger equation [40]. Its approximations include adiabatic or Born-Oppenheimer approximation for nuclei [58];

independent valence electron approximation in an average effective potential collectively determined by the nuclei and all electrons with the exchange-correlation energy approximated as a local functional of the electron density either in terms of the local spin density approximation (LSDA) [59,60] or GGA; and replacement of the strong Coulomb potential of the nucleus and the tightly bound core electrons by a pseudopotential as an effective potential acting on valence electrons such as the ultrasoft pseudopotentials and the projector augmented wave (PAW) method [61].

The electron density is solved iteratively until the energy of the ground-state configuration converges. Its Helmholtz energy at finite temperature includes the contributions from thermal electrons and phonons in terms of the quasiharmonic approximation (QHA) as follows [21]

$$F = E^0 + F^{el} + F^{vib} = E - TS \qquad \textit{Eq. 10}$$

$$E = E^0 + E^{el} + E^{vib} \qquad \textit{Eq. 11}$$

$$S = S^{el} + S^{vib} \qquad \textit{Eq. 12}$$

where $F$, $E$, and $S$ are the Helmholtz energy, internal energy (the same as $U$ used in above thermodynamics), and entropy of the configuration, $F^{el}$, $E^{el}$, and $S^{el}$ the contributions of thermal electron to $F$, $E$, and $S$ based on the Fermi–Dirac statistics for electrons, and $F^{vib}$, $E^{vib}$, and $S^{vib}$ the vibrational contributions to $F$, $E$, and $S$ based on the Bose–Einstein statistics for phonons, respectively. The thermal electronic contributions can be obtained from the finite temperature generalization of ground-state energy of an interacting inhomogeneous electron gas by Mermin [20] as shown by Kohn and Sham [9]. The vibrational contributions can be obtained by QHA phonon calculations or Debye model [21,62] through the high throughput DFT Tool Kits (DFTTK) [63,64]. The details can be found in the literature [21,57]. For vibration-induced dipole-dipole long-range interactions, a mixed-space approach was developed by the author's group using supercells in the real space to account the short-range interactions, the analytical solution for the origin in the reciprocal space to represent the infinite in the real space, and an interpolation scheme between them [65–67].

An important development in the exchange-correlation energy approximation has emerged recently. In GGA [10–12] developed by Perdew and co-workers the exchange-correlation energy is treated as a functional of both the local electron density and its gradient, resulting in more accurate predictions than LSDA. The latest SCAN meta-GGA [13,14] introduces symmetry breaking for some systems regarded as strongly correlated with significantly improved the quantitatively correct ground-state energetics [17,19]. The further developed $r^2$SCAN are with improved accuracy, numerical stability, and efficiency [14–18]. The key discovery in the SCAN meta-GGA is that "certain strong correlations present as fluctuations in the exact symmetry-unbroken ground-state wavefunction are 'frozen' in symmetry-broken electron densities or spin densities of approximate DFT" [19]. Consequently, an approximate density functional with symmetry breaking, though less accurate than an exact functional, is more revealing with its utility demonstrated for a number of cases [17–19].

It seems plausible that those symmetry-broken electron densities may be sampled by phonon calculations of the ground-state configuration such as the negative thermal expansions predicted at low temperatures in ice and Si [68]. This symmetry-broken configurations are in a scale finer than the symmetry-broken configurations in the zentropy theory to be discussed in Section 4, with the latter derived from the internal degrees of freedom of the ground-state configuration.

Some typical internal degrees of freedom in condensed matter include magnetic pin polarization, electric dipole polarization, atomic short-range ordering, and various defects such as vacancy, dislocation, stacking fault, twin, grain boundary, voids, and other ones at even larger scale. The Helmholtz energies of the metastable non-ground-state configurations can be predicted using Eq. 10 to Eq. 12. For unstable non-ground-state configurations, their Helmholtz energies cannot be directly predicted due to imaginary frequencies in their phonon dispersion curves. On the other hand, they only appear as transitory states at finite temperature with their properties predicted by the statistical mixtures of stable and metastable configurations by the zentropy theory discussed in Section 4.

### 2.3 Statistical mechanics

Statistical mechanics introduced by Gibbs [5] considerers the probabilities of all *independent* configurations rather than the individual particles in the system that are commonly used in the literature. Consequently, each configuration is under the same constraints as the system as defined by Gibbs [5] and discussed at the beginning of the introduction of the present paper. Based on the combined law of thermodynamics [1] and differentiating the internal and external variables in the system and its surroundings, Gibbs [5] evaluated the entropy due to the distribution of various configuration in the system though without considering the entropy of each configuration itself.

Landau and Lifshitz [7] formulated quantum statistical mechanics by introducing the number of quantum states in terms of the energy interval of the mean fluctuation of energy of the system and obtained the entropy of the system in terms of the tracer of each quantum state. In the limit of the classical theory, they showed the entropy of a system as Gibbs did as follows

$$S = -k_B \sum_{k=1}^{m} p^k ln p^k \qquad \textit{Eq. 13}$$

where $m$ is the number of configurations, and $p^k$ the probability of configuration $k$. Consequently, for a system of canonical ensemble under constant NVT, the partition function of the system ($Z$) in relation to $F$ of the system and the partition function ($Z^k$) and internal energy ($E^k$) of each configuration are written as follows

$$Z = e^{-\frac{F}{k_B T}} = \sum_{k=1}^{m} Z^k = \sum_{k=1}^{m} e^{-\frac{E^k}{k_B T}} \qquad \textit{Eq. 14}$$

$$F = -k_B T ln Z + k_B T \left( \sum_{k=1}^{m} p^k ln Z^k - \sum_{k=1}^{m} p^k ln Z^k \right)$$
$$= \sum_{k=1}^{m} p^k E^k + k_B T \sum_{k=1}^{m} p^k ln p^k = \sum_{k=1}^{m} p^k E^k - TS \qquad \textit{Eq. 15}$$

$$p^k = \frac{Z^k}{Z} = \frac{e^{-\frac{E^k}{k_B T}}}{Z} = e^{-\frac{E^k - F}{k_B T}} \qquad \textit{Eq. 16}$$

It is noted that the addition and subtraction of the same quantity in the parenthesis of Eq. 15 facilitates the derivation of the last part of Eq. 15 and the connection between Eq. 13. and Eq. 15.

For a hypothetical system with only one configuration, Eq. 15 thus becomes

$$F = E^k \qquad \textit{Eq. 17}$$

Since $F = E^k - TS^k$ by definition, Eq. 17 gives $S^k = 0$ as $T \neq 0$, indicating that the configurations do not have any internal degrees of freedom, i.e., they are all pure quantum configurations. In DFT developed after the quantum statistical mechanics [7], the ground-state configuration is not a pure quantum configuration, and its entropy is calculated by Eq. 12. The formula of the partition function thus needs to be modified as shown in Section 4, i.e., the zentropy theory.

# 3 Hillert nonequilibrium thermodynamics, Onsager theorem, Prigogine entropy balance, and Agren atomic mobility

## 3.1 Hillert nonequilibrium thermodynamics

As mentioned in Section 2.1, Gibbs [2] used Clausius' definition of entropy exchange ($d_Q S$) due to reversible heat exchange between the system and its surroundings to derive Eq. 1. The exchange of mass between the system and its surroundings brings both the exchange of internal energy and entropy. Furthermore, irreversible internal processes inside the system produce entropy based on the 2$^{nd}$ law of thermodynamics. The total entropy change of the system is thus written as follows [22,33,57]

$$dS = \frac{dQ}{T} + \sum_{i=1}^{c} S_i dN_i + d_{ip}S \qquad \text{Eq. 18}$$

where $S_i$ is the partial entropy of component *i* defined as

$$S_i = \left(\frac{\partial S}{\partial N_i}\right)_{dQ=0, d_{ip}S=0, N_{j\neq i}} \qquad \text{Eq. 19}$$

The work exchange between the system and its surroundings does not enter Eq. 2 directly, but indirectly by affecting internal processes. It is important to note that the total entropy change contains the contributions from both exchanges between the system and its surroundings (first two terms in Eq. 18) and the internal processes (last term). Therefore, $dS$ can be either positive or negative, which is not in contradiction with the 2nd law of thermodynamics as the 2nd law of thermodynamics concerns only the entropy production due to an independent internal process represented by Eq. 2.

Consequently, the combined law of thermodynamics with internal processes can be revised from Eq. 1 as follows [22,33,57]

$$dU = dQ + dW + \sum_{i=1}^{c} U_i dN_i = TdS + dW + \sum_{i=1}^{c} \mu_i dN_i - Td_{ip}S \qquad \textit{Eq. 20}$$

$$\mu_i = U_i - TS_i = \left(\frac{\partial U}{\partial N_i}\right)_{dS=0, dW=0, d_{ip}S=0, X^a \neq N_i} \qquad \textit{Eq. 21}$$

where $dW$ represent all types of work including mechanical, electric, and magnetic work, and $U_i$ denotes the partial internal energy of component $i$ defined as follows [45,57],

$$U_i = \left(\frac{\partial U}{\partial N_i}\right)_{dQ=0, dW=0, N_{j \neq i}} \qquad \textit{Eq. 22}$$

The chemical potential is defined by Eq. 21 with contributions from both partial internal energy and partial entropy of the component [45,57]. It should be emphasized that the constraints for the partial internal energy of Eq. 21 are different from those of Eq. 22 as $dS = 0$ requires the heat exchange between the system and its surroundings when $dN_i$ is exchanged between the system and its surroundings as shown by Eq. 18, while $dQ = 0$ is for an adiabatic system.

The explicit inclusion of $Td_{ip}S$ in Eq. 20 was emphasized by Hillert [22]. To differentiate Gibbs equilibrium thermodynamics, i.e., Eq. 3, the author proposes to name Eq. 20 "*Hillert nonequilibrium thermodynamics*" in contrast to Gibbs equilibrium thermodynamics or simply "*Hillert thermodynamics*" vs Gibbs thermodynamics. As mentioned in Section 2.1, one may thus define the *Hillert energy* by $\Phi = G - \sum_{i=1}^{c} \mu_i N_i$ and re-write the combined law of thermodynamics as follows

$$dG = dU - d(TS) + d(PV) = -SdT + VdP + \sum_{i=1}^{c} N_i d\mu_i - Td_{ip}S \qquad Eq.\ 23$$

$$d\Phi = dG - \sum_{i=1}^{c} \mu_i dN_i - \sum_{i=1}^{c} N_i d\mu_i = -SdT + VdP - \sum_{i=1}^{c} N_i d\mu_i - Td_{ip}S \qquad Eq.\ 24$$

Consequently, Eq. 20 and Eq. 24 may be termed as Hillert nonequilibrium thermodynamics in contrast to Gibbs equilibrium thermodynamics represented by Eq. 1 and Eq. 3.

For systems with multiple internal processes, the entropy production can be written as [22,57]

$$d_{ip}S = \frac{1}{T}\sum_{j=1}^{m} D_j d\xi_j \qquad Eq.\ 25$$

where $D_j$ and $\xi_j$ are a pair of internal conjugate variables for the $j^{th}$ internal process. However, Hillert [22] followed Onsager theorem by taking into account the possibility that those internal processes may interact with each other though he explicitly stated that "this is usually called ***phenomenological equation*** because it is not based on any physical model". As to be discussed in Section 5, the internal variables of $D_j$ and $\xi_j$ represent *independent* internal processes and are in analogy to the external variables of $Y^a$ and $X^a$ in Eq. 3. Eq. 20 can thus be re-written as follows

$$dU = \sum_{a=1}^{n} Y^a dX^a - \sum_{j=1}^{m} D_j d\xi_j \qquad Eq.\ 26$$

$$D_j d\xi_j \geq 0 \qquad Eq.\ 27$$

It should be remembered that the entropy change, $dS$ in the first summation in Eq. 26 now includes the second summation in Eq. 26. Furthermore, $d\xi_j$ may include more than one $dX^a$ such as chemical reactions that Hillert discussed [22] where $d\xi_j$ represents the amount of the products formed, and $D_j$ is the weighted chemical potentials of products minus those of reactants so that the combined internal process is an independent one, with more details shown in Section 3.1.

Eq. 27 depicts that each internal process must result in positive entropy production [57], while Hillert [22] stated that "the second law is derived only for the whole system", which is less restrictive. This difference is due to whether the internal processes are independent of each other or not. In Eq. 26, each pair of $Y^a$ and $X^a$ is independent of each other, so is each pair of $D_j$ and $\xi_j$. If not, they should be combined and be represented by one new pair of $D_j'$ and $\xi_j'$, just like the chemical reactions discussed by Hillert [22], resulting in $D_j' d\xi_j' \geq 0$. In analogy to Eq. 18, the entropy production of each independent internal process can be divided into four actions: (1) heat generation $\left(d_{ip}Q\right)$, (2) consumption of some components as reactants $\left(dN_{r,j}\right)$, (3) production of some components as products $\left(dN_{p,k}\right)$, and (4) reorganization of its configurations $\left(d_{ip}S^{config}\right)$, as follows [32,44],

$$d_{ip}S = \frac{d_{ip}Q}{T} - \sum\nolimits_j S_j dN_{r,j} + \sum\nolimits_k S_k dN_{p,k} + d_{ip}S^{config} = \frac{D}{T} d\xi \qquad \textit{Eq. 28}$$

where $S_j$ and $S_k$ are the partial entropies of reactant $j$ and product $k$ of the internal process and can be used to describe internal chemical reactions [69].

As Hillert discussed [22], the driving force for a chemical reaction can be written as follows

$$D = -\left(\sum\nolimits_k \mu_k \frac{\partial N_{p,k}}{\partial \xi} - \sum\nolimits_j \mu_j \frac{\partial N_{r,k}}{\partial \xi}\right) > 0 \qquad \textit{Eq. 29}$$

It is evident that chemical potentials of some components will increase and those of other components will decrease during a chemical reaction, but these individual changes are not independent, and their combination results in an independent internal process and a positive entropy production and positive driving force as shown by Eq. 28 and Eq. 29, respectively.

The last part of Eq. 28 is based on the *linear proportionality* approximation. In reality, internal processes do not obey linear proportionality in general, and one can select a small enough $d\xi$ to neglect the higher-order terms and perform integrations along the pathway for overall entropy production. To study the stability of an internal process or a system such as the limit of instability and critical points, one must include higher-order terms beyond linear proportionality as discussed in detail by Hillert [22]. Eq. 20 or Eq. 26 depicts that the internal energy is a function of all $X^a$ and internal variables of $\xi$, i.e., $U(X^a, \xi)$, so are all the potentials as the derivatives of $U(X^a, \xi)$ with respect to its natural variables. This relationship enables the development of the theory of cross phenomena in Section 5.

### 3.2 Onsager theorem

Experimental observations show that the transport of one molar quantity can be driven by the gradients of both its conjugate potentials and non-conjugate potentials. Based on the thermoelectric phenomena where the electric current can be driven by both electric field and temperature gradient and vice versa and the *conjecture* by Thomson in 1854 that the cross coefficients are equal, Onsager [24] proposed the phenomenological relations that any flux is proportional to all independent driving forces in the system as follows

$$J_{\xi_j}^{Onsager} = -\sum_k L_{\xi_j \xi_k} \nabla D_k \qquad \textit{Eq. 30}$$

where $L_{\xi_j \xi_k}$ is the kinetic coefficient for the flux of molar quantity $\xi_j$ due to $\nabla D_k = D_k/\Delta z$ with the summation going over gradients of all potentials, where $\Delta z$ is the distance of the unit transport process. Eq. 30 indicates that the flux of $\xi_j$ depends on all driving forces, in an *apparent* accordance with experimental observations. The rate of entropy production, $d_{ip}\dot{S}$, is then written as follows

$$\frac{T}{V} d_{ip}\dot{S} = -J_{\xi_j}^{Onsager} \nabla D_j = \nabla D_j \sum_{\xi_k} L_{\xi_j \xi_k} \nabla D_k \qquad \textit{Eq. 31}$$

From "the assumption of microscopic reversibility" [24], Onsager articulated that the phenomenological kinetic coefficient matrix in Eq. 30 is symmetric, i.e.,

$$L_{\xi_j \xi_k} = L_{\xi_k \xi_j} \qquad \textit{Eq. 32}$$

It is important to note that the principle of microscopic reversibility is based on "the fluctuations in a system which has been left isolated for a length of time that is normally sufficient to secure thermodynamic equilibrium" [24], and Onsager himself pointed out that "the reversible fundamental laws of dynamics are not compatible with absolutely irreversible processes" [24].

It is thus evident that the Onsager theorem discussed above is not on a strong foundation and has been strongly criticized by Truesdell and co-workers [70,71], while Hillert [22] emphasized its ***phenomenological*** nature, and Balluffi et al. [23] mentioned that it is not as fundamental as the 1st and 2nd laws of thermodynamics which was also stated by Onsager himself [24] as quoted in the introduction of the present paper . The present author started to look into this during his sabbatical leave when worked on thermodiffusion with Murch and his team [50–54], i.e., atomic diffusion driven by temperature gradient. Through molecular dynamics (MD) simulations, one can evaluate the kinetic coefficients through time integrals of correlation functions involving the microscopic fluxes of atoms and heat using the Green–Kubo formulae [50–54]. However, it is not straightforward to cast these coefficients into the flux equations as the driving forces and the frame of references are not explicitly spelled out, though the common practice is to treat them as Onsager phenomenological kinetic coefficients, which results in two coefficients for one component in a binary system as evidenced by Eq. 30. On the other hand, it is well-known that there is only one diffusivity for each diffusion component, namely its tracer diffusivity or atomic mobility in the lattice-fixed frame of reference [72]. This inconsistence was puzzling to the author. After seven years of investigations, the author developed the theory of cross phenomena by examining the inconsistences in the Onsager theorem and deriving the flux equations from Hillert nonequilibrium thermodynamics as shown by Eq. 26, which is presented in Section 5.

### 3.3 Prigogine entropy balance

There are many efforts in further evaluating entropy production beyond Eq. 31 in most irreversible thermodynamics textbooks such as by de Groot and Mazur[73] and Kondepudi and Prigogine [74]. Unfortunately, all of them were based on the Gibbs equilibrium thermodynamics by re-writing Eq. 1 or Eq. 3 as follows, without realizing $d_{ip}S = 0$

$$dS = \frac{1}{T}dU - \frac{1}{T}dW - \sum_{i=1}^{c}\frac{\mu_i}{T}dN_i \qquad \text{Eq. 33}$$

For example, Kondepudi and Prigogine [74] considered a system without work, and Eq. 33 becomes

$$dS = \frac{1}{T}dU - \sum_{i=1}^{c} \frac{\mu_i}{T} dN_i = d_e S \qquad \textit{Eq. 34}$$

where the last portion was added in the present work with $d_e S$ being the equilibrium entropy change or entropy current between the system and its surroundings [74]. Its time-dependent form can be written as

$$\frac{dS}{dt} = \dot{S} = \frac{1}{T}\dot{U} - \sum_{i=1}^{c} \frac{\mu_i}{T} \dot{N}_i = {}_e\dot{S} \qquad \textit{Eq. 35}$$

As discussed in detail in a recent publication by the present author [33], Kondepudi and Prigogine [74] obtained the following equation by re-using the 1st law of thermodynamics

$$\dot{S} = {}_e\dot{S} = {}_e\dot{S} + {}_{ip}\dot{S} \qquad \textit{Eq. 36}$$

where ${}_{ip}\dot{S}$ is the entropy production rate due to irreversible internal processes, denoted by $\sigma$ in ref. [74]. This incorrect outcome is due to the circular use of the 1st law of thermodynamics as discussed by the present author [33] because Eq. 34 is for ${}_{ip}\dot{S} = 0$ as defined in Gibbs equilibrium thermodynamics.

It is important to realize that the 1st law of thermodynamics concerns only the exchanges of heat, work, and mass between the system and its surroundings as shown by first part of Eq. 20. The entropy change is based on Eq. 18 which shows that $d_{ip}S$ is part of $dS$ and thus must be subtracted as shown in the second part of Eq. 20. Consequently, Eq. 33 and Eq. 34 are valid only with $d_{ip}S = 0$, i.e., no internal processes in the system for entropy production. In the extended irreversible thermodynamics [75,76] that includes fluxes as basic independent variables, Eq. 20 can be re-arranged as follows

$$\dot{S} = \frac{1}{T}\dot{U} - \frac{1}{T}\dot{W} - \sum_{i=1}^{c} \frac{\mu_i}{T}\dot{N_i} + {}_{ip}\dot{S} = {}_{e}\dot{S} + {}_{ip}\dot{S} \qquad Eq.\ 37$$

with ${}_{ip}\dot{S}$ related to the divergency of fluxes [75,76]. It remains how to define the flux equations.

### 3.4 Agren atomic mobility

In 1980s, the author learned from Agren [77] that each diffusion component has one diffusion mobility in the lattice-fixed frame of reference ($M_i$), which can be related to its tracer diffusivity ($D_i^*$) or kinetic coefficient ($L_i$) in the linear relationship between its flux and chemical potential gradient. When this chemical potential gradient is changed to the concentration gradients of all independent diffusion components, one obtains a vector of intrinsic diffusivity in the lattice-fixed frame of reference (${}^{i}D_{ik} = L_i \frac{\partial \mu_i}{\partial c_k}$) from the dependence of chemical potential on compositions of all components ($c_k$). When the lattice-fixed frame of reference is changed to the volume-fixed frame of reference while keeping the chemical potential gradient, one obtains a kinetic coefficient vector ($L'_{ij}$) due to the dependence of volume on all components. When both changes are made, one obtains the chemical or interdiffusion diffusivity vector ($D_{ik} = \sum_j L'_{ij} \frac{\partial \mu_j}{\partial c_k}$).

Andersson and Ågren [72] presented an elegant discussion among all the diffusion coefficients and their computational implementation. The relationships among these kinetic coefficients are shown in Figure 1. As discussed by the present author [33], it is thus clear that both the intrinsic and chemical diffusivity coefficients are related to the mobility and thermodynamic factors defined by the derivatives between chemical potentials and compositions ($\frac{\partial \mu_j}{\partial c_k}$) and are not independent kinetic coefficients as shown by the Maxwell-Stefan diffusion equation [78,79].

*Figure 1: Relationships among tracer diffusivity, atomic mobility, kinetic L parameters, and intrinsic and chemical diffusivities* [32].

Many atomic mobility databases have been developed with Gibbs energy functions modeled by the CALPHAD method [80,81] and used to simulate diffusional processes in multicomponent alloys, including joining of dissimilar materials [82–84]. One interesting scenario is when one component diffuses much faster than other components and thus takes much short time to reach its chemical equilibrium, such as carbon (C) in steels. Since the chemical potential of C is significantly affected by other elements such as Si, the concentration of C can become more inhomogeneous than its initial concentration profile, resulting in the migration of C from low concentration to high concentration regions, i.e. so-called uphill diffusion, postulated and observed by Darken [85–87]. The uphill diffusion in Fe-Si-C alloys with inhomogeneous Si concentrations are successfully simulated by means of the mobility database and Dictra software developed by Agren and his team [80,88] and discussed in detail by the present author [33]. It is important to point out that C always diffuses from high chemical potential regions to low chemical potential regions as its chemical potential is affected by the Si concentration which takes much longer time to be homogenized.

It should be mentioned that uphill diffusion also occurs when the solution is unstable with respect to composition fluctuations with $\frac{\partial \mu_i}{\partial c_i} < 0$, commonly referred to as spinodal decomposition [89–92]. Even though the atomic mobility is always positive, the intrinsic or chemical diffusivity is negative inside a spinodal the solution due to $\frac{\partial \mu_i}{\partial c_i} < 0$. The instability-driven spinodal decompositions play a central role in the formation of patterns [74,91].

One important phenomenon in diffusion is the Kirkendall effect [93–95] due to the difference in fluxes of substitutional elements which results in a net vacancy flow, which has also been successfully simulated using the atomic mobility databases together with thermodynamic databases [96,97]. The flux of vacancy, $J_{Va}$, can be written as follows

$$J_{Va} = -\sum_{i \in S} J_i \qquad \textit{Eq. 38}$$

where $i \in S$ denotes all substitutional diffusion components, and $J_i$ the flux of component $i$. The unbalanced fluxes of substitutional components can result in a net vacancy flow in accordance with the vacancy-mediate diffusion mechanism.

The atomic mobility databases have been mostly evaluated from the experimentally measured chemical diffusivity in combination with CALPHAD thermodynamic databases. Th author's group developed a fully DFT-based approach to accurately predict the tracer diffusivity for fcc [98–102], bcc [103], and hcp [103–106] phases using the classic transition state theory (TST) [107,108], the nudged elastic band (NEB) method [109], and the five frequency model [110]. A similar approach was developed for prediction of interstitial diffusion coefficients in the literature at the same time [111].

Using the thermodynamic and mobility databases, Höglund and Ågren [112] simulated C diffusion in steel driven by a temperature gradient under the framework of Onsager theorem by treating the coefficient in front of the temperature gradient as a constant. As it will be shown in Section 5.4.2, this coefficient is the product of atomic mobility and the derivative of chemical potential with respect to temperature and depends on both temperature and composition [33].

# 4 Zentropy theory

## 4.1 Multiscale entropy and its coarse-graining

In statistical mechanics, Eq. 13 represents the entropy among the configurations and does not include entropy of each configuration. If the entropies of individual configurations are not zero, their contributions to the total entropy of the system need to be added to the total entropy of the system. Consequently, the total entropy of the system is the sum of Gibbs entropy among configurations in terms of Eq. 13 plus the statistical average of the entropy of each configuration, $S^k$, as follows [44]

$$S = \sum_{k=1}^{m} p^k S^k - k_B \sum_{k=1}^{m} p^k ln p^k = \int_0^T \frac{C_P}{T} dT \qquad Eq.\ 39$$

The last part in Eq. 39 represents the experimentally measured total entropy through the integration of heat capacity using the 3rd law of thermodynamics stipulating $S = 0$ at $T = 0\ K$, commonly referred as Clausius entropy in the literature. Eq. 39 shows that the Gibbs entropy and Clausius entropy are equal only when $S^k = 0$ for each configuration, i.e., all configurations are pure quantum configuration as discussed in relation to Eq. 17.

In principle, one can continuously dive into finer scale configurations until the pure quantum configuration is reached, which is intractable in today's computing capability due to too many configurations. One option is to stop at the scale of the ground-state configuration of the system based on the state-of-the-art DFT-based quantum mechanics as discussed in Section 2.2 and explore all non-ground-state symmetry-broken excited configurations with respect to the internal degrees of freedom of the ground-state configuration. The entropy and free energy of each configuration can then be predicted by Eq. 10 to Eq. 12, respectively.

The usefulness of Eq. 39 relies on following two premises:

- The supercell size in DFT calculations based on today's computing capability is large enough to capture the electron and phonon interactions in the system so that each configuration can be treated as independent.
- The configurations are ergodic so that there are no unrepresented configurations when the configurations are thermally fluctuated in the system under the given external constraints.

These two premises are interconnected as the larger the supercell size, the more the number of configurations. In principle, convergency tests are needed in terms of the predicted properties of the system. Alternatively, comparison with experimental observations, if available, can be used as a guidance in order to save computing expenses.

In connecting information entropy and internal processes, the present author and his collaborators [44] pointed out that the second summation in Eq. 39 is the same as the information entropy at the scale of observation discussed by Szilard [113,114] and Shannon [115,116]. The difference between the Clausius entropy, also often referred as thermodynamic entropy, and the information entropy is thus the first summation in Eq. 39, i.e., the sum of probability-weighted entropy of each configuration. Similar discussions were presented in the literature such as recently by Zivieri [117] by considering the conversion of the thermodynamic entropy (last term in Eq. 39) into information entropy (second summation in Eq. 39) at the expense of negentropy (the first summation in Eq. 39), with the negentropy or "negative entropy" corresponding to the loss of thermodynamic entropy from the magnetic skyrmion which are topological swirling spin textures of individual configurations employed as information carriers, i.e.,

$$S^{info} = -k_B \sum_{k=1}^{m} p^k lnp^k = \int_0^T \frac{C_P}{T} dT - \sum_{k=1}^{m} p^k S^k \qquad \textit{Eq. 40}$$

In condensed matter physics, many phenomena are due to short-range interactions and can be predicted within today's computing capability [118–120]. For disordered liquid phases, Hong and the present author proposed to represent liquid configurations in terms of the coordination number of each atom and accurately predicted the liquid entropies and melting temperatures of a number of pure elements, compounds, and oxides [121]. An alternative approach is to develop machine learning (ML) models from fewer DFT calculations and use them to predict properties of large supercells and a large number of configurations [122,123].

### 4.2 Statistical mechanics based on the zentropy theory

Applying Eq. 39 to a system of canonical ensemble under constant NVT, Eq. 14 to Eq. 16 are revised as follows

$$F = \sum_{k=1}^{m} p^k E^k - TS = \sum_{k=1}^{m} p^k E^k - T\left(\sum_{k=1}^{m} p^k S^k - k_B \sum_{k=1}^{m} p^k lnp^k\right)$$

$$= \sum_{k=1}^{m} p^k (E^k - TS^k) + k_B T \sum_{k=1}^{m} p^k lnp^k \qquad \textit{Eq. 41}$$

$$= \sum_{k=1}^{m} p^k F^k + k_B T \sum_{k=1}^{m} p^k lnp^k$$

$$Z = e^{-\frac{F}{k_B T}} = \sum_{k=1}^{m} Z^k = \sum_{k=1}^{m} e^{-\frac{F^k}{k_B T}} \qquad \textit{Eq. 42}$$

$$p^k = \frac{Z^k}{Z} = e^{-\frac{F^k - F}{k_B T}} \qquad \textit{Eq. 43}$$

As it can be seen, $E^k$ in Eq. 14 to Eq. 16 is replaced by $F^k$ in Eq. 41 to Eq. 43, and they are identical when $S^k = 0$. Eq. 39 to Eq. 43 was recently suggested to be termed as zentropy theory [49], and its significances are discussed below.

For condensed matter, Eq. 39 integrates three scientific domains, i.e., quantum mechanics, statistical mechanics, and thermodynamics in terms of first summation, second summation, and the integration, which have been largely separated from each other. Efforts have been made to bridge the gaps between them through bottom-up approaches by considering the thermal electronic and phonon distributions of the ground-state configurations or using effective Hamiltonian followed by MD or Monte Carlo (MC) simulations, plus limited work on *ab initio* molecular dynamic (AIMD) and quantum Monte Carlo (QMC) simulations. However, quantitative agreement between predictions and experiments is lacking in the literature due to the intrinsic limitations of existing approaches as discussed in our recent publication [48], i.e., the simultaneous considerations of all internal degrees of freedom as a function of external constraints.

In materials science, the system of interest is at the microstructure level, consisting of individual phases and their temporal and spatial evolutions. One can take each individual phase as a subsystem of investigation and predict its properties by means of the zentropy theory by defining its configurations in terms of atomic, magnetic, electrical, and defect configurations and calculating their entropies and free energies through DFT. These properties can then be used in phase-field methods to simulate their temporal and spatial evolutions [91]. Another important property needed for phase-field simulations is the interfacial energy between phases and grains, and its accurate prediction is still lacking with similar reason mentioned in the previous paragraph [48]. One potential solution is to fit the effective Hamiltonian to free energies of all configurations predicted by the zentropy theory followed by MD or MC simulations. While this can account properly the first summation in Eq. 39 for interfaces, and additional steps are needed in treating the second summation in Eq. 39 so the total entropy can be accurately predicted. The new method developed by Hong and the present author seems promising in addressing the disordering in interfaces [121].

## 4.3 Prediction of emergent behaviors in magnetic materials by zentropy theory

### *4.3.1 General discussion of emergent behaviors*

In the present work, emergent behaviors refer to abnormal responses of the system to external potentials, i.e., temperature, pressure, stress, and electric and magnetic fields and are typically related to $1^{st}$, $2^{nd}$, and $3^{rd}$ derivatives of free energy of the system. With the free energy of condensed matter accurately predicted by the zentropy theory, it is anticipated that their emergent behaviors can be predicted accordingly. It can be seen from Eq. 39 and Eq. 41 that the nonlinear or emergent behaviors primarily originate from the logarithmic term, i.e., the statistical competition among configurations with respect to external stimuli.

Gibbs thermodynamics, i.e., Eq. 3, shows that the 1st derivative of energy with respect to a molar quantity gives its conjugate potential with other molar quantities as natural variables kept constant, i.e.,

$$Y^a = \frac{\partial U}{\partial X^a} \qquad \textit{Eq. 44}$$

By defining a free energy, such as Helmholtz energy, the 1st derivative of Helmholtz energy with respect to temperature gives the negative of its conjugate molar quantity, entropy, i.e.,

$$S = -\frac{\partial F}{\partial T} \qquad \textit{Eq. 45}$$

The deviation of entropy from that of the ground-state configuration in terms of QHA phonon calculations is considered to be anharmonic [124].

The 2nd derivatives of energy are defined by the 1st derivatives between molar quantities and potentials and are listed in Table 1, and between molar quantities and between potentials in Table 2. Both tables are symmetric through the change of variable sequency in the 2nd derivatives of free energy. The diagonal quantities in Table 1 are derivatives between conjugate variables and are positive for a stable system as shown by Eq. 9, while the off-diagonal quantities are derivatives between non-conjugate variables and can be negative such as negative thermal expansion in INVAR alloys [125–127]. The derivative between conjugate molar quantity and potential diverges positively at the limit of stability, such as $\frac{\partial S}{\partial T} = +\infty$ shown in Section 2.1, resulting the maximum anharmonicity. While the derivatives between non-conjugate molar quantity and potential can also diverge negatively at the limit of stability, i.e.,

$$\frac{\partial X^a}{\partial Y^b} = \frac{\partial X^b}{\partial Y^a} = \pm\infty \qquad \textit{Eq. 46}$$

Most of the quantities in Table 1 are well known properties, while those in the last row and last column are related to chemical reactions where the amount of a component changes with respect to potentials as discussed by Prigogine and co-workers [74]. The author assigned some names to the quantities in the last row, but the names for those in the last column remain to be assigned.

*Table 1: Physical quantities related to 1st directives between molar quantities (first column) and potentials (first row), slightly modified from Ref.* [33].

*Table 2: Cross phenomenon coefficients represented by 1st derivatives between potentials, slightly modified from Ref.* [33].

The derivatives between potentials in Table 2 are less discussed in the literature though Gibbs [3,4] presented a number of them in connection with equilibria involving solid and interfaces between phases such as $\frac{\partial \mu_i}{\partial T}$, $\frac{\partial \mu_i}{\partial P}$, and $\frac{\partial P}{\partial T}$, which were termed as cross phenomena in the literature where the Onsager theorem was based on [24,25]. As a matter of fact, all off-diagonal quantities in Table 1 and Table 2 represent cross phenomena between non-conjugate variables and will be further discussed in terms of the theory of cross phenomena in Section 5.

### *4.3.2 Emergent behaviors in magnetic materials*

Magnetic materials have been used in developing the zentropy theory starting with anti-invar Ce [42,43] due to its importance in Mg alloys [128,129], followed by INVAR $Fe_3Pt$ due to its negative thermal expansion [46,68,130,131]. Their configurations are defined by various magnetic spin configurations. The zentropy theory has been used to predict the magnetic transitions in bcc-Fe[132], fcc-Ni[133], orthorhombic $BaFe_2As_2$ [134–136], and $YNiO_3$ [48]. An attempt was made recently to apply the zentropy theory to the ferroelectric transition in $PbTiO_3$ with encouraging results [137,138].

The ground-state configuration of Ce is nonmagnetic (NM), and its room temperature phase is ferromagnetic (FM). We started with these two configurations with a one-atom supercell [42] and found out that a mean-field magnetic spin flipping term in free energy was needed in order to obtain the critical point in its T-P phase diagram as commonly done in the literature [139]. To consider the spin-flipping magnetic (SFM) configurations and magnetic spin domain walls due to the co-existence of the NM and FM configuration, an antiferromagnetic (AFM) configuration was added with a two-atom supercell [43]. With the DFT-predicted Helmholtz energies of these three configurations, the critical point and associated anomalies were accurately predicted without the mean-field term [43,45,68].

The critical point defines the boundary between the 1$^{st}$- and 2$^{nd}$-order transitions. For the 1$^{st}$-order transition, the two-phase equilibrium is determined by the minimization of the system Helmholtz energy with the low-temperature phase having higher probability of the ground-state configuration than the high-temperature phase. At the critical point, the two phases merge into one phase. In Ce, the predicted critical point is with $T_{cp} = 546\ K$ and $P_{cp} = 2.05\ GPa$ and in good accordance with the range of experimental observations. It is observed that the probability of the ground-state configuration is about 0.5, i.e., $p^g = 0.5$, which has been used to define the 2$^{nd}$-order transition temperature and may also be rationalized to represent that the ground-state configuration is no longer dominant and loses its percolation in the system.

The predicted T-P and T-V phase diagrams show remarkable agreement with experimental observations without any models and fitting parameters [43,49,68]. The positive divergency of thermal expansion was predicted at the critical point, attributing to the larger volumes of the AFM and FM non-ground-state configurations than that of the NM ground-state configuration. This can be seen from the following volume and thermal expansion equations of a system based on the zentropy theory [140]

$$V = \frac{\partial G}{\partial P} = \sum_{k=1}^{m} p^k V^k = V^g + \sum_{k=1}^{m} p^k (V^k - V^g) \qquad \textit{Eq. 47}$$

$$\frac{\partial V}{\partial T} = \sum_{k=1}^{m} \left[ p^k \frac{\partial V^k}{\partial T} + \frac{\partial p^k}{\partial T} (V^k - V^g) \right] \qquad \textit{Eq. 48}$$

where $V^g$ and $V^k$ are the volumes of the ground-state configuration and the non-ground-state configuration $k$, respectively. Eq. 47 shows that $V^k > V^g$ results in $V > V^g$ due to $p^k \geq 0$. With $\frac{\partial p^g}{\partial T} < 0$ for the ground-state configuration and $\frac{\partial p^k}{\partial T} > 0$ for non-ground-state configurations, $\frac{\partial V}{\partial T} > 0$, results in the positive divergence at the critical point, i.e., $\frac{\partial V}{\partial T} = +\infty$, as shown in the T-V phase diagram of Ce [49,68].

It is immediately evident that if $V^k < V^g$, it is possible that $V < V^g$ and $\frac{\partial V}{\partial T} < 0$ under certain T and P ranges. This is indeed the case for $Fe_3Pt$ where all SFM non-ground-state configurations have smaller volumes than the FM ground-state configuration [46]. Using a 12-atom supercell with $2^9 = 512$ collinear SFM configurations, of which 37 is unique due to symmetry, the T-P and T-V phase diagrams of $Fe_3Pt$ were accurately predicted using the zentropy theory including the critical point, the negative divergence of thermal expansion at the critical point, and the negative thermal expansion in a range of T and P combinations, showing remarkable agreement with available experimental observations without fitting parameters [46,68,130,131].

As in the case of Ce, $p^g = 0.5$ is also observed at the critical point in $Fe_3Pt$ with $T_{cp} = 141\ K$ and $P_{cp} = 5.81\ GPa$ within the range of experimental observations [46]. As mentioned above, , it is convenient to define a 2nd-order transition by $p^g = 0.5$. On the other hand, the definition of a 2nd-order transition is by the discontinuity of the 2nd-order derivative of Helmholtz energy to T, i.e., heat capacity. In Ce, it was found that the maximum of the sum of the magnetic plus electronic heat capacities is at 500 K under 2.05 GPa, lower than 546 K with $p^g = 0.5$ at the same pressure [43]. Fr bcc-Fe, the 2nd-order transition temperature is in the descending order of $p^g = 0.5$, the maximum of the total $C_p$, and the maximum of the magnetic heat capacity with the middle one in accordance with the agreed value in the literature from experimental measurements [46]. For $Fe_3Pt$, our recent analysis showed similar results obtained from the peaks of magnetic heat capacity or the change of entropy short-range ordering [131]. It is noted that the predicted heat capacity shows a maximum, but not a discontinuous jump with respect to temperature as shown by experimental observations, probably due to the small supercell size used in the DFT calculations.

In $BaFe_2As_2$ [134–136], the ground-state configuration is a spin density wave (SDW) AFM configuration. Within the *ab* plane of an orthorhombic structure a stripe like Fe spin ordering pattern is formed with antiparallel nearest-neighbor Fe spins along the *c* axis. The predicted SDW ordering temperature with $p^g = 0.5$ as a function of pressure using a 40 atom supercell, shows remarkable agreement with experimental measurements [136]. Furthermore, the temperature with $p^g = 0.9999$ as a function of pressure was discovered to be in accordance with a characteristic superconducting-like temperature, which is in alignment with our postulations of superconductivity [32] though the definition of superconducting configuration remains elusive and is part of our on-going investigations [141,142].

#### *4.3.3 On $YNiO_3$ with strongly correlated physics*

$YNiO_3$ in the $RNiO_3$ family is with strongly correlated physics and commonly investigated by using a combination of DFT and dynamical mean-field theory calculations in the literature [143,144]. The present author's team recently applied the zentropy theory to predict the AFM-to-PM transition in $YNiO_3$ [48]. The ground-state configuration of $YNiO_3$ is rather complex with half of the Ni atoms having a negligible magnetic moment and the other half arranged in an AFM configuration [47]. Without magnetic spins, its minimum cell would contain four $ABO_3$ formula units with 20 atoms. For magnetic spin configurations, a minimum cell of 16 formula units with 80 atoms is needed, resulting in 35 possible supercells from the spatial arrangements of four 20-atom base cells. Considering 8 Ni atoms with zero and 8 Ni atoms with nonzero magnetic moments creates 256 spin configurations for each supercell, and 37 of them are symmetry-independent for the 80-atom 1×2×2 supercell.

With the Helmholtz energies of the 37 configurations predicted by DFT calculations, their probabilities are obtained using the zentropy theory and plotted in Figure 2 (i) using $F^k$ and (ii) using $E^k$ for the partition function of each configuration, respectively. The predicted AFM-PM transition temperature of 144 K with $p^g = 0.5$ using $F^k$ agrees remarkably with the experimentally measured 145 K, while the one using $E^{\mathrm{k}}$ results in a temperature of 81 K with $p^g = 0.5$. This difference is further shown in the $T - P$ phase diagram in Figure 2 (iii). This demonstrates the critical importance to include the phonon contributions to each configuration and replace the total energy in partition function of each configuration by its Helmholtz energy. Figure 2 (iv) plots the temperature dependency of the magnetic spin short-range ordering (SRO) under ambient pressure, showing that the AFM-to-PM transition is near the temperature where SRO is between 0.4 and 0.5. Here SRO is defined as follows

$$SRO = 1 - \frac{SD(T)}{SD(SQS)} \qquad \textit{Eq. 49}$$

where $SD(T)$ and $D(SQS)$ are the standard deviations of the distribution of total magnetic moments within the 1$^{st}$ coordination sphere of Ni atom from the zentropy theory and by the spin-special quasirandom model for mimicking the high-temperature paramagnetic state [145,146].

It is thus demonstrated that the strongly correlated physics in $YNiO_3$ can be predicted by the zentropy theory through the consideration of statistical mixture of the ground-state and non-ground-state configurations. It reveals the important role of spatial fluctuations, derived from the competition among spin ground-state and nonground-state configurations, in the AFM-to-PM phase transition. Therefore, the strong correlations in $YNiO_3$ reside in each configuration and are exemplified by the spin and phonon interactions, while all independent configurations are coupled through statistical mechanics. It is thus critical to include the spin and phonon contributions in each configuration in statistical mechanics in order to fully capture those strong correlated physics as shown by the zentropy theory.

*Figure 2: $YNiO_3$ (i) predicted probabilities of the ground-state (red symbols) and non-ground-state (their sum in open symbols) configurations as a function of temperature under ambient pressure; (ii) same as (i) but with $E^k$ instead of $F^k$ for the partition function of configuration $k$; (iii) predicted T-P phase diagram superimposed with the AFM-to-PM transition temperature measured by experiments (red dot) and predicted using $E^k$ (blue dot); (iv) predicted SRO as a function of temperature under ambient pressure* [48].

### *4.3.4 Miscalculation of entropy on "microscopic violation of second law of thermodynamics"*

There have been many discussions on the violation of 2nd law of thermodynamics in the literature since the 2nd law of thermodynamics was formulated. The latest argument was on the violation in the microscopic scale in terms of the fluctuation theorem [147,148]. Many investigations were performed aiming to demonstrate the existence of trajectories, with statistically significant probability, between two states of a system with the work done ($\Delta W$) smaller than the equilibrium Helmholtz energy difference between the two states ($\Delta F$) [149–153].

The Helmholtz energy difference between the two local equilibrium states represents the minimum amount of heat or work to move the system between these two states through reversible processes, so that the less amount of work observed indicates the violation of the 2nd law of thermodynamics. This means that a closed circle from one state to another state in a system by doing work $\Delta W$ and then back to the original state by releasing the energy $\Delta F$ would result in a net energy production, i.e., $(|\Delta F| - |\Delta W|) > 0$, which could be extracted from the system. This was demonstrated by Maillet et al. [152] who constructed a single-electron transistor two-level system with $\Delta F = 0$. By coupling a single thermodynamic trajectory level to a single heat bath, they were able to extract the amount of work up to large fractions of $k_B T$. However, Maillet et al. [152] pointed out that the requirement of an external intervention makes the 2nd law of thermodynamics remain valid.

More generally, the central question is how to fully count the entropy change for internal processes. By considering an internal process as a system, the change of entropy between two moments and the total entropy at any moment are represented by Eq. 28 and Eq. 39, respectively. Using the reversible Brownian motion as an example, the author [32] pointed out that as soon as the probability for an atom to jump over to the next vacant site becomes statistically significant, additional entropy is introduced due to the 2nd summation in Eq. 39. The driving force for this internal reversible process is the thermal energy taken from its surroundings so Eq. 28 becomes

$$d_{ip}S = \frac{d_{ip}Q}{T} + d_{ip}S^{config} = 0 \qquad \textit{Eq. 50}$$

with $d_{ip}Q < 0$ and $Td_{ip}S^{config} = -d_{ip}Q$. The increase of $d_{ip}S^{config}$ is represented by the possibilities that the atom moves either forward or backward. When the atom makes either move, the thermal energy is reversibly given back to its surroundings, leaving the entropies of both the internal process as a system and its surroundings unchanged.

In most discussions related to the "violation of second law of thermodynamics", the 2nd summation in Eq. 39 is not considered, resulting in lower accounting of entropy production represented by $d_{ip}S^{config}$ so Eq. 50 becomes negative, thus a false conclusion on the violation of the 2nd law of thermodynamics.

# 5 Theory of cross phenomena

## 5.1 Issues on Onsager theorem and Prigogine entropy balance

Today's irreversible thermodynamics is largely based on the Onsager theorem and Prigogine entropy balance. As discussed in Section 3.3, Prigogine entropy balance is based on Gibbs equilibrium thermodynamics and can thus not be used to describe irreversible processes. Furthermore, the irreversible thermodynamics established by Prigogine and co-workers [74] are based on Onsager theorem, which has several internal inconsistences as discussed by the present author [32,33] and shown below in the present paper.

Through over seven years of study since 2014, the present author formulated four fundamental questions and comments concerning the Onsager theorem as follows [32,33]

1. As a symmetric matrix can be diagonalized to obtain its eigen values (kinetic coefficients) and the eigen vector (the set of independent driving forces), what is the eigen vector after the diagonalization of Onsager flux equations? Since eigenvectors and eigenvalues play important roles in all the areas where linear algebra is applied such as flux discussed in the present work, they need to be defined rigorously and should be governed by Hillert nonequilibrium thermodynamics as presented in Section 3.1 to go beyond phenomenological representation of experimental observations.
2. When $D_j = 0$, $d\xi_j$ may not be zero because the Onsager flux equation relates $d\xi_j$ to all driving forces. Does this mean that the internal processes are not independent? If the internal processes cannot be separated into independent ones such as chemical reactions that require the collective actions of all reactants, those dependent processes need to be combined into one independent process for it to be included as a term in Hillert nonequilibrium thermodynamics as discussed in Section 3.1 for chemical reactions.

3. The entropy production for the internal process in question 2 is zero as shown by Eq. 31, implying that a non-zero $J_{\xi_j}^{Onsager}$ does not produce entropy. Is this in conflict with the 2nd law of thermodynamics? This is an extension of the question 2 above. Based on the 2nd law of thermodynamics, any independent internal process must result in a positive entropy production as discussed in Section 3.1 and represented by Eq. 27. If an internal process results in non or negative entropy production, it is not an independent process, but part of a larger independent internal process, and needs to be combined into the independent internal process as discussed above.
4. If the microscopic reversibility and Gibbs thermodynamics hold locally, so does the Gibbs-Duhem equation as shown by Eq. 6, signifying that not all potentials or their gradients could be varied independently. Does this mean at least one of the driving forces in Onsager flux equation must be a molar quantity? If so, which one? From the stability criteria shown by Eq. 9, one immediately realizes that it should be the molar quantity of which the flux is considered because they change in the same direction and appear together in the combined law of thermodynamics.

Phenomenological oriented approaches aim to correlate experimental observations with possible causes through the top-down macroscopic view of the responses of a system to the observer's actions. For a more complete understanding of the responses, one must examine both the temporal and spatial resolutions of the observations and consider the internal processes in the system and the interactions between the system and its surroundings by microscopically eliminating the degrees of freedom and appropriately averaging the whole responses. The limited spatial resolution is very much similar to the parable of the blind men and the elephant articulated by Perdew et al. [19] that each individual observes a different portion of a complex system and thus is incomplete by themselves alone. This holistic view enabled Perdew and his collaborators [13–19] to develop the SCAN meta-GGA with significantly improved ground-state energetics as briefly discussed in Section 2.2. The limited temporal resolution is excellently illustrated by the illusion of thaumatrope where when a disk with a picture on each side is spined fast enough, the two pictures merge into one such as bird in the cage [154].

This illusion was demonstrated by the transition of $PbTiO_3$ from the ferroelectric tetragonal structure at low temperature to the paraelectric cubic structure at high temperature observed by X-ray and neutron diffractions. However, the ferroelectric tetragonal structure persists locally at both low and high temperatures as observed through the XAFS (x-ray-absorption fine structure) analysis with the time and spatial resolutions being $\sim 10^{-16}$ sec and 1$^{st}$ to 4$^{th}$ nearest neighbor shells [155–157] and AIMD simulations [158] with the overall lattice parameters from the x-ray diffractions [159]. The faster switching of the ferroelectric tetragonal structure among different orientations at high temperature results in the macroscopic cubic structure observed by the x-ray diffractions with lower temporal and spatial resolutions as shown in the video from the AIMD simulations [137]. This raises a similar question concerning the noncollinear magnetic spin configurations, i.e., how can one be sure that the observation is not an illusion?

As discussed below, the phenomenological Onsager flux equations originate from the dependence of a potential not only on its own conjugate molar quantity, but also on all other natural variables as mentioned in Section 3.1. Consequently, when one natural variable, whether it is a molar quantity or potential, is changed, the potentials of all other molar quantities are affected, resulting in driving forces for all other molar quantities to transport from their high potential regions to their respective low potential regions. Phenomenological correlations between the fluxes of those molar quantities to the change of the initial natural variable are similar to the illusions mentioned above. This is in analogy to a Chinese idiom: Move one hair and move the whole body.

## 5.2 Formulation of internal processes and entropy production

The compact form of Hillert nonequilibrium thermodynamics is presented by Eq. 26 with the first summation denoting the interactions between the system and its surroundings and the second summation representing the *independent* internal processes. It is important to emphasize again that the entropy change in the first summation includes the total entropy production of all internal processes inside the system depicted by the second summation. For dependent internal processes such as chemical reactions discussed by Hillert [22], Kondepudi and Prigogine [74], and in Section 3.1, they should be combined to establish an independent internal process when using the combined law of thermodynamics.

Eq. 26 shows only terms of products between conjugate variables without any cross-terms between non-conjugate variables. It is known that the equilibrium of a system is reached when each potential has the same value everywhere in the system. If this is not the case, the difference of the potential provides the driving force for its conjugate molar quantity to migrate from high potential regions to low potential regions, resulting in a flux of the molar quantity in the system. This flux results in the change of other potentials in the system and provides driving forces for their molar quantities to migrate. Therefore, to answer the first question to Onsager theorem presented in Section 5.1 above, the eigen vector of driving forces is the gradient of all independent potentials shown by Eq. 26.

From the above discussion, the rate of the entropy production per volume due to the $j^{th}$ internal process in a sufficiently small region with a thickness of Δz and area of $A$ can be written as

$$\frac{Td_{ip}\dot{S}_j}{V} = \frac{D_j d\dot{\xi}_j}{A\Delta z} = -\frac{d\dot{\xi}_j}{A}\frac{\Delta Y_{\xi_j}}{\Delta z} = -\frac{d\dot{\xi}_j}{A}\nabla Y_{\xi_j} \qquad \textit{Eq. 51}$$

with $D_j = -\Delta Y_{\xi_j}$ and $Y_{\xi_j}$ being the conjugate potential of the internal variable of $\xi_j$, and $\nabla Y_{\xi_j}$ the gradient of $Y_{\xi_j}$. The flux of $\xi_j$, $J_{\xi_j}$, can then be defined as follows with the driving force being the gradient of its conjugate potential

$$J_{\xi_j} = \frac{d\dot{\xi}_j}{A} = -L_{\xi_j}\nabla Y_{\xi_j} \qquad \textit{Eq. 52}$$

where $L_{\xi_j}$ is the kinetic coefficient for the change of $\xi_j$. Eq. 51 can be re-written as

$$\frac{Td_{ip}\dot{S}_j}{V} = L_{\xi_j}\left(\nabla Y_{\xi_j}\right)^2 \qquad \textit{Eq. 53}$$

It can be seen that Eq. 52 addresses both first and second questions to Onsager theorem presented in Section 5.1 above, i.e., the eigen vector being the potential gradients and a zero flux of a molar quantity with a zero gradient of its conjugate potential. Eq. 53 answers the third question, i.e., zero entropy production for an internal process with a zero gradient of its potential.

The answer to the fourth question is more complex. As discussed by Hillert [22], there are two types of stable equilibria with zero entropy productions with Eq. 27 written as

$$D_j d\xi_j = 0 \qquad \textit{Eq. 54}$$

with either $D_j = 0$ or $d\xi_j = 0$. Hillert [22] termed the latter as equilibrium under freezing-in conditions when the internal process could not take place due to its high kinetic barrier in comparison to the thermal energy, and it is commonly referred as metastable equilibrium as the system free energy could be further reduced through some internal processes, such as diamond vs graphite.

For a homogeneous metastable system, the internal variable $\xi_j$ becomes an independent variable of the system, and all the properties of the system are thus dependent on $\xi_j$ as discussed by Hillert [22] and mentioned in Section 3.1. The Gibbs-Duhem equation, Eq. 6 or more precisely Eq. 24 or Eq. 26 with $Td_{ip}S = \sum_{j=1}^{m} D_j d\xi_j = 0$, is thus applicable to such a homogeneous system, i.e., the potentials in the system are not independent to each other. However, when there are internal processes inside the system $Td_{ip}S > 0$, Eq. 24 applies, and the Gibbs-Duhem equation is no longer valid.

Nevertheless, the free energy at each moment in time and space in the system with well-defined internal variables can be evaluated and used to calculate the properties through its derivatives, while the change of free energy also depends on the change of the internal variables which are affected by all internal processes as shown by Eq. 26, resulting the cross phenomena that Onsager theorem aims to represent and are discussed in the next section.

### 5.3 Formulation of theory of cross-phenomena

As discussed above, each potential is a function of its conjugate molar quantity and all other natural variables of the system. Consequently, the gradient of a potential can be written in terms of the gradients of its conjugate molar quantity and all other natural variables with some of them being molar quantities and some of them being potentials as follows

$$\nabla Y_{\xi_j} = \frac{\partial Y_{\xi_j}}{\partial \xi_j}\nabla \xi_j + \sum\nolimits_{\xi_k \neq \xi_j} \frac{\partial Y_{\xi_j}}{\partial Y_{\xi_k}}\nabla Y_{\xi_k} + \sum\nolimits_{\xi_l \neq \xi_j, \xi_k} \frac{\partial Y_{\xi_j}}{\partial \xi_l}\nabla \xi_l \qquad \textit{Eq. 55}$$

where the first and second summations represent potential and molar quantity natural variables, respectively. In different experimental settings, one or more natural variables are controlled, but the summations in Eq. 55 must include all natural variables because their values will be affected by the changes of other natural variables internally. Those natural variables can be either all molar quantities or all potentials excluding $Y_{\xi_j}$ and $\xi_j$.

The flux equation, Eq. 52, can thus be further expanded as follows

$$J_{\xi_j} = -L_{\xi_j}\nabla Y_{\xi_j} = -L_{\xi_j}\left(\frac{\partial Y_{\xi_j}}{\partial \xi_j}\nabla \xi_j + \sum\nolimits_{\xi_k \neq \xi_j} \frac{\partial Y_{\xi_j}}{\partial Y_{\xi_k}}\nabla Y_{\xi_k} + \sum\nolimits_{\xi_l \neq \xi_j, \xi_k} \frac{\partial Y_{\xi_j}}{\partial \xi_l}\nabla \xi_l\right) \qquad \textit{Eq. 56}$$

It is noted that $L_{\xi_j}$ would also depend on all independent variables in Eq. 56, i.e. $\xi_j$, $Y_{\xi_k}$, and $\xi_l$. The two summations in Eq. 56 represent the cross phenomena in the system. Eq. 56 is termed as the theory of cross phenomena [32], and its significances are as follows

1. The flow of a molar quantity is ***solely*** driven by the gradient of its conjugate potential with a characteristic kinetic coefficient under the linear proportionality approximation, derived from the Hillert nonequilibrium thermodynamics ***without phenomenological*** considerations.
2. Both the potential gradient and the characteristic kinetic coefficient are functions of all independent variables in an internal process, resulting in the cross-phenomena shown by the two summations in Eq. 56.
3. The product of the flux of a molar quantity and its conjugate potential results in the entropy production rate due to the internal process that contributes to the energy change rate of the system as one term in Hillert nonequilibrium thermodynamics.

The detailed applications of cross phenomena were discussed by the present author in the literature [32], including thermoelectricity, thermodiffusion, chemical interdiffusion, electromigration, electrocaloric effect, and electromechanical effect. They are briefly reviewed and updated in following sections.

### 5.4 Applications of theory of cross phenomena

There are four common types of transport phenomena in condensed matter physics: heat, electron, mass, and fluid, commonly represented by the Fourier's, Ohm's, Fick's, and Darcy's laws based on experimental observations, with $\xi_j = S$, $c_e$, $c_i$ and $V$, and $\nabla Y_{\xi_j} = T$, $E$, $\mu_i$, and $-P$, respectively. In the Fourier's, Ohm's, and Darcy's laws, Eq. 52 is used with the gradient of conjugate potentials as the driving force, so their linear proportionality coefficients represent the kinetic coefficients, i.e., $L_{\xi_j}$ in Eq. 52. On the other hand, the concentration gradients, i.e. $\nabla c_i$, are used in Fick's law in terms of Eq. 56, which are not the true driving forces for diffusion.

Therefore, the kinetic coefficients $L_{\xi_j}$:s represent the direct relation between conjugate variables and discussed in detail by the present author [32]. The coefficients for cross-phenomena are the products of the kinetic coefficients and the thermodynamic properties represented by derivatives in Eq. 56. The derivatives between two potentials are relatively easy to measure as temperature, pressure, and electrical field are typically controlled experimentally, while the derivative between molar quantities can be predicted computationally [22,45,57]. This represents an integration of complimentary experimental and computational strengths connected by the Maxwell relation as follows

$$\frac{\partial Y^a}{\partial X^b} = \frac{\partial^2 \Phi}{\partial X^b \partial X^a} = \frac{\partial Y^b}{\partial X^a} \qquad \textit{Eq. 57}$$

$$\frac{\partial Y^a}{\partial Y^b} = \frac{\partial^2 \Phi}{\partial Y^b \partial X^a} = -\frac{\partial X^b}{\partial X^a} \qquad \textit{Eq. 58}$$

### *5.4.1 Thermoelectricity*

Thermoelectricity concerns the conduction of electrons or holes due to an externally applied temperature gradient. For electron and entropy conductions, the flux equations are written as follows

$$J_e = -L_e \nabla \mu_e \qquad \textit{Eq. 59}$$

$$J_S = -L_S \nabla T \qquad \textit{Eq. 60}$$

where $L_e$ and $L_S$ are the electrical and thermal conductivity, and $\nabla\mu_e$ and $\nabla T$ the gradients of chemical potential of electrons and temperature. Based on Hillert thermodynamics, the conjugate molar quantity of temperature is entropy rather than commonly used heat, and their relation is denoted by Eq. 18.

Let us conduct a virtual experiment to investigate the performance of a thermoelectric material as follows

- Initial condition at $t = 0$ is with $\nabla T = 0$, $\nabla c_e = 0$, $\nabla\mu_e = 0$.
- At a very short time with $t = \varepsilon$, a temperature gradient is applied, resulting in $\nabla T > 0$. Assuming no electron migration yet, one has $\nabla c_e = 0$. However, $\nabla\mu_e \neq 0$ due to the dependence of $\mu_e$ on both electron concentration and temperature, $\mu_e(c_e, T)$.
- At $t > \varepsilon$: nonuniform $\mu_e$ induces electron migration and results in a concentration gradient of electron in the thermoelectric material: $\nabla c_e \neq 0$.
- For open-circuit experimental setting, electrons remain in the system. For time long enough, the system reaches a steady state with $J_e = 0$ and $\nabla\mu_e = 0$, balanced by $\nabla c_e$ and $\nabla T$.
- The electron concentration profile in the system induces an internal electric field, and its voltage, $\nabla V_e$, can be measured.
- The ratio of voltage to $\nabla T$ is determined and termed as Seebeck coefficient

$$S_{T,e} = \frac{\nabla V_e}{\nabla T} \qquad \textit{Eq. 61}$$

Re-writing Eq. 59 in terms of Eq. 56, one obtains

$$J_e = -L_e \nabla \mu_e = -L_e \left( \frac{\partial \mu_e}{\partial c_e} \nabla c_e + \frac{\partial \mu_e}{\partial T} \nabla T \right) = -L_e (\Phi_{ee} \nabla c_e - S_e \nabla T) \qquad \textit{Eq. 62}$$

where $\Phi_{ee}$ and $S_e$ are the thermodynamic factor and partial entropy of electrons. Under the steady state condition with $J_e = 0$, Eq. 62 gives

$$S_e = -\frac{\nabla V_e}{\nabla T} \qquad \textit{Eq. 63}$$

The combination of Eq. 61 and Eq. 63 results in the Seebeck coefficient

$$S_{T,e} = \frac{\nabla V_e}{\nabla T} = -S_e \qquad \textit{Eq. 64}$$

It is thus shown that the Seebeck coefficient for electron conduction, i.e., n-type of thermoelectric materials with electrons added to the conduction band, is negative of partial entropy of electrons.

For p-type thermoelectric materials with positively charged holes added to the valence band, one has under steady state condition,

$$J_h = -L_h (\nabla V_h - S_h \nabla T) = 0 \qquad \textit{Eq. 65}$$

$$S_{T,h} = \frac{\nabla V_h}{\nabla T} = S_h \qquad \textit{Eq. 66}$$

where $L_h$, $\nabla V_h$, $S_h$, and $S_{T,h}$ are the kinetic coefficient, voltage, partial entropy, and Seebeck coefficient of holes in p-type thermoelectric materials, respectively. One thus has positive Seebeck coefficients for p-type thermoelectric materials. With the Helmholtz energy of electrons predicted by the DFT calculations (see Eq. 10 and Eq. 12 ), the author's group used the theory of cross phenomena to accurately predict the Seebeck coefficients for several n- and p-type thermoelectric materials [160,161].

Furthermore, the migrating electrons carry entropy with them, inducing an entropy current as follows

$$J_S = TS_eJ_e = -TS_eL_e\nabla\mu_e \qquad \text{Eq. 67}$$

This entropy current results in the Peltier effect with the Peltier coefficient defined by the division of entropy current to the electrical current as follows, noting that the electrical current is in the opposite direction of the electron flux

$$\Pi = \frac{J_S}{-J_e} = -TS_e = TS_{T,e} \qquad \text{Eq. 68}$$

The Thomson relation between the Peltier and Seebeck coefficients comes out automatically

$$\Pi = TS_{T,e} \qquad \text{Eq. 69}$$

The above discussion demonstrates that both Peltier and Seebeck coefficients are thermodynamic quantities related to the derivative of chemical potential of electron to temperature and equal to the partial entropy of electrons through the Maxwell relation as shown in Table 2 and as follows

$$S_{T,e} = \frac{\partial\mu_e}{\partial T} = \frac{\partial^2 G}{\partial T\partial c_e} = -\frac{\partial S}{\partial c_e} = -S_e \qquad \text{Eq. 70}$$

### *5.4.2 Thermodiffusion*

Thermodiffusion is similar to thermoelectricity by changing the electrons to atoms and can be studied using the same virtual experiment presented in Section 5.4.1 above. The additional complexity is the dependence of the chemical potential on the concentrations of all components and the change of volume with respect to temperature, both affect atomic migration significantly. due to the second summation in Eq. 56. Consequently, the flux equation of thermodiffusion is written as

$$J_i = -L_i \nabla \mu_i = -L_i \left( \sum_j \Phi_{ij} \nabla c_j - P_i \nabla V - S_i \nabla T \right) \qquad \textit{Eq. 71}$$

For non-ideal solutions, $L_i$, $\Phi_{ij} = \frac{\partial \mu_i}{\partial c_j}$, $P_i = -\frac{\partial \mu_i}{\partial V} = \frac{\partial P}{\partial c_i}$, and $S_i = -\frac{\partial \mu_i}{\partial T} = \frac{\partial S}{\partial c_i}$ can strongly depend on temperature and compositions of the solutions, resulting in a diffusion component switching their segregation regions as a function of composition and temperature [162–169] and also the Kirkendall effect as shown by Eq. 38 in Section 3.4.

For binary systems where most experiments are reported in the literature, Eq. 71 is written as follows

$$J_A = -L_A \nabla \mu_A = -L_A (\Phi_{AA} \nabla c_A + \Phi_{AB} \nabla c_B - P_A \nabla V - S_A \nabla T) \qquad \textit{Eq. 72}$$

$$J_B = -L_B \nabla \mu_B = -L_B (\Phi_{BA} \nabla c_A + \Phi_{BB} \nabla c_B - P_B \nabla V - S_B \nabla T) \qquad \textit{Eq. 73}$$

Under steady state conditions with $J_A = J_B = 0$, eliminating $\nabla c_A$ from the equations results in

$$(\Phi_{BB}\Phi_{AA} - \Phi_{AB}\Phi_{BA})\nabla c_B = (P_B\Phi_{AA} - P_A\Phi_{BA})\nabla V + (S_B\Phi_{AA} - S_A\Phi_{BA})\nabla T \qquad \textit{Eq. 74}$$

The Soret coefficient [170] is commonly defined by the negative ratio of the $\nabla c_B$ with respect to $\nabla T$ as follows

$$S_{T,B} = -\frac{\nabla c_B}{c_B \nabla T} = -\frac{1}{c_B} \frac{S_B\Phi_{AA} - S_A\Phi_{BA}}{\Phi_{BB}\Phi_{AA} - \Phi_{AB}\Phi_{BA}} \left( 1 + \frac{P_B\Phi_{AA} - P_A\Phi_{BA}}{S_B\Phi_{AA} - S_A\Phi_{BA}} \frac{\nabla V}{\nabla T} \right) \qquad \textit{Eq. 75}$$

Eq. 75 shows that $S_{T,B} = 0$ when the one of the following conditions is met

$$S_B\Phi_{AA} - S_A\Phi_{BA} = 0 \qquad \textit{Eq. 76}$$

$$S_B\Phi_{AA} - S_A\Phi_{BA} + (P_B\Phi_{AA} - P_A\Phi_{BA})\frac{\nabla V}{\nabla T} = 0 \qquad \textit{Eq. 77}$$

For systems with negative thermal expansion at certain composition and temperature ranges, it is possible that the Soret coefficient changes its sign with respect to composition and temperature as observed in the literature [45,49,130].

The entropy flow due to atomic diffusions can be obtained by generalizing Eq. 67 as follows

$$J_S = T\sum_i S_i J_i \quad \textit{Eq. 78}$$

This entropy flux contributes to the total entropy flux though very small due to the small flux in typical atomic diffusion.

While there are many discussions in the literature on Soret effects as reviewed by the present author [32], the actual quantitative simulations of thermodiffusion is rare due to the lack of thermodynamic and kinetic databases. One very interesting work was reported by Höglund and Ågren [112] who simulated the thermodiffusion of carbon in an Fe-32%Ni-0.14%C (weight percent, wt%) alloy using available thermodynamic and mobility databases as discussed in Section 3.4. As discussed in detail by the present author [32], it seems that their simulations can be further improved by connecting the commonly used heat of transport for component $i$, $Q_i^*$, to its partial entropy as shown in Eq. 71, i.e.,

$$Q_i^* = T\frac{\partial \mu_i}{\partial T} = -TS_i \quad \textit{Eq. 79}$$

### *5.4.3 Electromigration*

Electromigration concerns the atomic diffusion driven by an external electrical field and is the most serious reliability issue in interconnect metallization and flip chip solder joints in electronic devices [171–174]. The initial internal process in electromigration is the electrical current, followed by entropy conduction, internal stress, and atomic diffusion. The flux for electromigration can be written as follows based on Eq. 56

$$J_i = -L_i \nabla \mu_i = -L_i \left( \sum_j \Phi_{ij} \nabla c_j - S_i \nabla T - \boldsymbol{\varepsilon}_i \nabla \boldsymbol{\sigma} - \boldsymbol{\theta}_i \nabla \boldsymbol{E} \right) \qquad \text{Eq. 80}$$

where partial strain $\boldsymbol{\varepsilon}_i$ and partial electrical displacement $\boldsymbol{\theta}_i$ are listed in Table 2. The Kirkendall effect shown by Eq. 38 is more profound in electromigration as the electric field heavily affects migration directions of atoms with some models even including a source term for the nonequilibrium vacancy concentration [175,176].

For electromigration of pure metals with approximations of $\theta_i \approx 0$, $\nabla c_j \approx 0$, $\nabla T \approx 0$ and $\nabla \boldsymbol{\sigma} \approx 0$. With the internal electric field represented by the electron concentration as discussed in Section 5.4.1, Eq. 80 can be approximated as follows in analogy to Eq. 62

$$J_A = -L_A \nabla \mu_A = -L_A \frac{\partial \mu_A}{\partial c_e} \nabla c_e = -L_A \Phi_{Ae} \nabla c_e \qquad \text{Eq. 81}$$

Therefore, pure metals diffuses in the same direction of the decrease of electron concentration, i.e. the direction of electron flow, from cathode to anode, and the vacancy diffuses in the opposite direction ($J_{Va} = -J_A$, see Eq. 38) resulting in the formation of voids on the cathode side [171]. The entropy flow can be calculated by Eq. 80 with the summation including both the element and electron.

### *5.4.4 Electrocaloric effect*

The electrocaloric effect (ECE) concerns the cross phenomenon between entropy conduction due to electron flow and an external electrical field for heating or cooling [177–183]. When an external electrical field is applied to the system, an electric current is generated due to the Ohm's law, and the electrons carry entropy with them, resulting in aconduction in terms of Eq. 67 due to the electric current and Eq. 57 due to the temperature gradient. Considering only the flows of entropy and electrons, the entropy flux can be obtained from the general form of Eq. 56 as follows

$$J_S = -L_S \nabla T = -L_S \left( \frac{\partial T}{\partial S} \nabla S + \frac{\partial T}{\partial E} \nabla E \right) = -L_S \left( \frac{T}{C_P} \nabla S - \frac{1}{S_\theta} \nabla E \right) \qquad \textit{Eq. 82}$$

The derivatives in Eq. 82 are related to heat capacity and partial entropy with respect to electric displacement as depicted in Table 1 and Table 2, respectively.

Furthermore, the derivatives in Eq. 82 represent the direct and indirect methods used in the literature to characterize the ECE materials. In the direct method, the temperature increase under the *adiabatic condition* as a function of electric field is measured, i.e. $(\Delta T / \Delta E)_{\Delta Q=0}$. The adiabatic condition with $\Delta Q = 0$ is identical to the isentropic condition with $\Delta S = 0$ for a closed equilibrium system. The temperature change can be obtained through integration of the second derivative in Eq. 82 as follows

$$\Delta T_{direct} = \int_{E_1}^{E_2} \frac{\partial T}{\partial E} dE = -\int_{E_1}^{E_2} \frac{1}{S_\theta} dE \qquad \textit{Eq. 83}$$

In the indirect method, the heat production under the *isothermal condition* as a function of electrical field is measured, i.e. $(\Delta Q / \Delta E)_T$ with $\Delta Q$ treated to be the same as $T\Delta S$. The entropy change in the indirect method can be calculated through integration and is then used to calculate the anticipated temperature increase using the inverse of the first derivative in Eq. 82 or heat capacity of the materials as follows

$$\Delta S_{indirect} = \int_{E_1}^{E_2} \frac{\partial S}{\partial E} dE = \int_{T_1}^{T_1+\Delta T_{indirect}} \frac{\partial S}{\partial T} dT = \int_{T_1}^{T_1+\Delta T_{indirect}} \frac{C_P}{T} dT \qquad \textit{Eq. 84}$$

It is interesting to note that the results from both methods contribute to the entropy flow per Eq. 82. Therefore, maximizing both $\Delta T_{direct}$ and $\Delta S_{indirect}$ can improve the performance of ECE materials. Furthermore, while $\Delta T_{direct}$ and $\Delta T_{indirect}$ can both be used to characterize the performance of ECE materials, they are likely different from each other since both methods start from the same state, i.e. $(T_1, S_1, E_1)$, but end at different states, i.e. $(T_1 + \Delta T_{direct},\ E_2)_{S_1}$ *vs* $(S_1 + \Delta S,\ E_2)_{T_1}$. Therefore, it is in general that $\Delta T_{indirect} \neq \Delta T_{direct}$. The heat capacity in Eq. 84 is often assumed to be constant in the literature, resulting in the following approximated equation

$$T_1 \Delta S_{indirect} = C_P \Delta T_{indirect} \qquad \textit{Eq. 85}$$

Since $C_P$ varies dramatically with temperature near morphotropic phase boundaries (MPBs) where ECE is mostly investigated, this approximation could introduce large errors in evaluating ECE performance of materials.

### *5.4.5 Electromechanical effect*

The electromechanical effect concerns the relations between electric field and elastic deformation [184–188] with giant effects often observed near MPBs [189–192]. The electric current in piezoelectricity is created due to electric charge accumulation in response to externally applied stress as follows

$$J_e = -L_e \nabla \mu_e = -L_e \left( \frac{\partial \mu_e}{\partial c_e} \nabla c_e + \frac{\partial \mu_e}{\partial \sigma} \nabla \sigma \right) = -L_e (\Phi_{ee} \nabla c_e - \varepsilon_e \nabla \sigma) \qquad \textit{Eq. 86}$$

where $\varepsilon_e = \frac{\partial \varepsilon}{\partial c_e} = -\frac{\partial \mu_e}{\partial \sigma}$ is the partial strain, included in Table 2. With zero electrical current, the charge redistribution results in a voltage as follows

$$\nabla V_e = \varepsilon_e \nabla \sigma \qquad \textit{Eq. 87}$$

On the other hand, an externally applied electric field induces elastic deformation in the converse piezoelectric effect. The mechanical equilibrium is reached by vanishing stress gradient as follows

$$\nabla \sigma = \frac{\partial \sigma}{\partial \varepsilon} \nabla \varepsilon + \frac{\partial \sigma}{\partial c_e} \nabla c_e + \frac{\partial \sigma}{\partial E} \nabla E = 0 \qquad \textit{Eq. 88}$$

The zero electric current is written as follows

$$\nabla \mu_e = \frac{\partial \mu_e}{\partial c_e} \nabla c_e + \frac{\partial \mu_e}{\partial \varepsilon} \nabla \varepsilon + \frac{\partial \mu_e}{\partial E} \nabla E = 0 \qquad \textit{Eq. 89}$$

The partial derivatives in Eq. 88 and Eq. 89 are all included in Table 1 and Table 2. The giant electromechanical effects originate from the anomalies of these partial derivatives and can be predicted by the zentropy theory.

## 6 Summary and outlooks

In the present work, three distinct scientific domains developed in last 150 years are briefly reviewed, i.e., and equilibrium and nonequilibrium thermodynamics, statistical mechanics, and quantum mechanics. They have been largely separated from each other due to the different principles that each is based on and the different aspects of a complex system that each represents. In Gibbs equilibrium thermodynamics the entropy is inserted into the combined law of thermodynamics for systems under equilibrium, and in Hillert nonequilibrium thermodynamics the entropy production due to internal processes is added for systems with internal processes. Statistical mechanics depicts a top-down view of a system by considering configurations that the system is composed of and their statistical distributions in the system through partition functions of the system and individual configurations. DFT is a practical solution of quantum mechanics and focuses on the electronic structures, energetics, and phonon properties of the ground-state configuration of the system, aiming to predict the system behaviors from bottom-up.

The zentropy theory developed by the present author's group integrates DFT and statistical mechanics through following concepts and procedures

- Postulate that the configurations of a system are composed of the ground-state configuration from DFT and the non-ground-state excited configurations from the internal degrees of freedom of the ground-state configuration.
- Examine all the configurations of the system in terms of multiplicity and stability and predict their Helmholtz energies as a function of internal and external constraints through DFT calculations.
- Define the entropy of the system as the statistical entropy among all configurations plus the weighted sum of the entropy of each configuration by its statistical probability.
- Implement the revised statistical mechanics formalism by using the Helmholtz energy for the partition function of each configuration instead of commonly used total energy.

- Minimize the Helmholtz energy of the system with respect to internal degrees of freedom to obtain its equilibrium state with respect to external constraints.

It is demonstrated that the zentropy theory accurately predicts Helmholtz energy of magnetic materials and all other properties derived from Helmholtz energy such as T-P and T-V phase diagrams with critical points, $1^{st}$- and $2^{nd}$-order magnetic transitions, anomaly in thermodynamic properties, and positive and negative divergencies of thermal expansion, showing remarkable agreement with experimental observations.

Through analysis of Hillert nonequilibrium thermodynamics, it is concluded that the flux of a molar quantity is proportional *only* to the gradient of its conjugate potential rather than the gradients of all potentials depicted by the phenomenological Onsager theorem. It is shown that the observed dependence of the flux on the gradients of other potentials is due to the dependences of the conjugate potential on the other potentials in the system. These dependences are represented by the derivatives between potentials and can be accurately predicted by the zentropy theory as demonstrated for Seebeck coefficients of n thermoelectric materials. It is shown that the closer a system to a critical point or phase boundary, the larger those derivatives and the responses of a system to external stimuli.

There remain many challenging and complex problems to test the zentropy theory and the theory of cross phenomena in either smaller or larger systems. For smaller systems, one challenge is on superconductivity. Based on the existing success of the zentropy theory, the present author presented several postulations on superconductivity with the focus on the search for the superconducting configuration (SCC) as the ground-state configuration of a superconductor [32] with ongoing research activities [141]. Recently, the present author's team identified the superconducting and normal conducting configurations in both conventional and unconventional superconductors in terms differential charge density and presented promising prediction of superconducting temperature of Al using the zentropy theory [142].

While for larger and more complex systems beyond the current DFT calculations, such as plants, organisms, forests, societies, planets, the solar system, black holes, galaxies, and superclusters, there exist many intermediate levels of configurations between their ground-state configurations and observables. One potential approach is to take the inputs from both bottom-up quantum mechanics and top-down observations and cast them into the nested formula of the zentropy theory to develop practical solutions [33,44].

Another potential direction is to improve the accuracy of MD simulations through the inclusion of configurational entropy among configurations in evaluating the total entropy of the system as briefly mentioned in Section 4.2 for interfacial energy. This is particularly interesting for predicting the properties of liquid or amorphous where the configurations are not well defined as in crystals. The present author and his collaborators are working on addressing this issue through prediction of melting using the zentropy theory, aiming to capture the total entropy and free energy of liquid more accurately [121].

## 7 Acknowledgements

The author feels privileged to work with all his current and former students and numerous collaborators over the years at Penn State and around the world as reflected by the co-author names in the references cited in this paper. The present review article covers research outcomes supported by multiple funding agencies over multiple years as reflected in acknowledgements of the references cited. The current supports include the Endowed Dorothy Pate Enright Professorship at the Pennsylvania State University, U.S. Department of Energy Grant No. DE-SC0023185, DE-AR0001435, DE-NE0008945, and DE-NE0009288, U.S. National Science Foundation Grant No. 2229690, 2226976, and 2050069, and Office of Naval Research Grant No. N00014-21-1-2608 and N00014-23-2721.

## 8 References

[1] Gibbs J W 1948 *The collected works of J. Willard Gibbs: Vol. I Thermodynamics* (New Haven: Yale University Press, Vol. 1)

[2] Gibbs J W 1873 Graphical methods in the thermodynamics of fluids *Trans. Connect. Acad. II* **April-May** 309–42

[3] Gibbs J W 1876 On the equilibrium of heterogeneous substances *Trans. Connect. Acad. III* **May** 108–248

[4] Gibbs J 1878 On the equilibrium of heterogeneous substances *Trans. Connect. Acad. III* **July** 343–524

[5] Gibbs J W 1948 *The collected works of J. Willard Gibbs: Vol. II Statistical Mechanics* (New Haven: Yale University Press, Vol. II)

[6] Schrödinger E 1926 An Undulatory Theory of the Mechanics of Atoms and Molecules *Phys. Rev.* **28** 1049–70

[7] Landau L D and Lifshitz E M 1970 *Statistical Physics* (Oxford, New York: Pergamon Press Ltd.)

[8] Hohenberg P and Kohn W 1964 Inhomogeneous electron gas *Phys. Rev. B* **136** B864–71

[9] Kohn W and Sham L J 1965 Self-Consistent Equations Including Exchange and Correlation Effects *Phys. Rev.* **140** A1133–8

[10] Langreth D C and Perdew J P 1980 Theory of nonuniform electronic systems. I. Analysis of the gradient approximation and a generalization that works *Phys. Rev. B* **21** 5469–93

[11] Perdew J P, Chevary J A, Vosko S H, Jackson K A, Pederson M R, Singh D J and Fiolhais C 1992 Atoms, molecules, solids, and surfaces: Applications of the generalized gradient approximation for exchange and correlation *Phys. Rev. B* **46** 6671–87

[12] Perdew J P and Wang Y 1992 Accurate and simple analytic representation of the electron-gas correlation energy *Phys. Rev. B* **45** 13244

[13] Sun J, Ruzsinszky A and Perdew J 2015 Strongly Constrained and Appropriately Normed Semilocal Density Functional *Phys. Rev. Lett.* **115** 036402

[14] Furness J W, Kaplan A D, Ning J, Perdew J P and Sun J 2020 Accurate and Numerically Efficient r2SCAN Meta-Generalized Gradient Approximation *J. Phys. Chem. Lett.* **11** 8208–15

[15] Grimme S, Hansen A, Ehlert S and Mewes J M 2021 R2SCAN-3c: A “swiss army knife” composite electronic-structure method *J. Chem. Phys.* **154** 64103

[16] Kothakonda M, Kaplan A D, Isaacs E B, Bartel C J, Furness J W, Ning J, Wolverton C, Perdew J P and Sun J 2023 Testing the r 2 SCAN Density Functional for the Thermodynamic Stability of Solids with and without a van der Waals Correction *ACS Mater. Au* **3** 102–11

[17] Perdew J P, Chowdhury S T U R, Shahi C, Kaplan A D, Song D and Bylaska E J 2023 Symmetry Breaking with the SCAN Density Functional Describes Strong Correlation in the Singlet Carbon Dimer *J. Phys. Chem. A* **127** 384–9

[18] Maniar R, Withanage K P K, Shahi C, Kaplan A D, Perdew J P and Pederson M R 2023 Symmetry Breaking and Self-Interaction Correction in the Chromium Atom and Dimer *J. Chem. Phys.* **submitted**

[19] Perdew J P, Ruzsinszky A, Sun J, Nepal N K and Kaplan A D 2021 Interpretations of ground-state symmetry breaking and strong correlation in wavefunction and density

functional theories *Proc. Natl. Acad. Sci. U. S. A.* **118** e2017850118
[20] Mermin N D 1965 Thermal Properties of the Inhomogeneous Electron Gas *Phys. Rev.* **137** A1441–3
[21] Wang Y, Liu Z K and Chen L-Q 2004 Thermodynamic properties of Al, Ni, NiAl, and Ni3Al from first-principles calculations *Acta Mater.* **52** 2665–71
[22] Hillert M 2007 *Phase Equilibria, Phase Diagrams and Phase Transformations* (Cambridge: Cambridge University Press)
[23] Balluffi R W, Allen S M and Carter W C 2005 *Kinetics of Materials* (John Wiley and Sons)
[24] Onsager L 1931 Reciprocal Relations in Irreversible Processes, I *Phys. Rev.* **37** 405–26
[25] Onsager L 1931 Reciprocal relations in irreversible processes. II *Phys. Rev.* **37** 2265–79
[26] Prigogine I, Outer P and Herbo C 1948 Affinity and Reaction Rate Close to Equilibrium *J. Phys. Colloid Chem.* **52** 321–31
[27] Prigogine I 1951 The Equilibrium Hypothesis in Chemical Kinetics. *J. Phys. Chem.* **55** 765–74
[28] Prigonine I and Résibois P 1961 On the kinetics of the approach to equilibrium *Physica* **27** 629–46
[29] Prigogine I and Nicolis G 1967 On Symmetry-Breaking Instabilities in Dissipative Systems *J. Chem. Phys.* **46** 3542–50
[30] Prigogine I 1975 Dissipative structures, dynamics and entropy *Int. J. Quantum Chem.* **9-S9** 443–56
[31] Prigogine I 1978 Time, structure, and fluctuations *Science* **201** 777–85
[32] Liu Z K 2022 Theory of cross phenomena and their coefficients beyond Onsager theorem *Mater. Res. Lett.* **10** 393–439
[33] Liu Z K 2023 Thermodynamics and its prediction and CALPHAD modeling: Review, state of the art, and perspectives *CALPHAD* **82** 102580
[34] Brandão F, Horodecki M, Ng N, Oppenheim J and Wehner S 2015 The second laws of quantum thermodynamics *Proc. Natl. Acad. Sci.* **112** 3275–9
[35] Melkikh A V. 2021 Can Quantum Correlations Lead to Violation of the Second Law of Thermodynamics? *Entropy* **23** 573
[36] Kosloff R 2013 Quantum Thermodynamics: A Dynamical Viewpoint *Entropy* **15** 2100–28
[37] Zivieri R 2023 Trends in the Second Law of Thermodynamics *Entropy* **25** 1321
[38] Wang Y, Curtarolo S, Jiang C, Arroyave R, Wang T, Ceder G, Chen L Q and Liu Z K 2004 Ab initio lattice stability in comparison with CALPHAD lattice stability *CALPHAD* **28** 79–90
[39] Liu Z K, Chen L-Q, Raghavan P, Du Q, Sofo J O, Langer S A and Wolverton C 2004 An integrated framework for multi-scale materials simulation and design *J. Comput. Mater. Des.* **11** 183–99
[40] Liu Z K 2009 First-Principles calculations and CALPHAD modeling of thermodynamics *J. Phase Equilibria Diffus.* **30** 517–34
[41] Kaufman L and Bernstein H 1970 *Computer Calculation of Phase Diagrams* (New York: Academic Press Inc.)
[42] Wang Y, Hector L G, Zhang H, Shang S L, Chen L Q and Liu Z K 2008 Thermodynamics of the Ce γ–α transition: Density-functional study *Phys. Rev. B* **78**

104113
[43] Wang Y, Hector Jr L G, Zhang H, Shang S L, Chen L Q and Liu Z K 2009 A thermodynamic framework for a system with itinerant-electron magnetism *J. Phys. Condens. Matter* **21** 326003
[44] Liu Z K, Li B and Lin H 2019 Multiscale Entropy and Its Implications to Critical Phenomena, Emergent Behaviors, and Information *J. Phase Equilibria Diffus.* **40** 508–21
[45] Liu Z K 2020 Computational thermodynamics and its applications *Acta Mater.* **200** 745–92
[46] Wang Y, Shang S L, Zhang H, Chen L-Q and Liu Z-K 2010 Thermodynamic fluctuations in magnetic states: Fe3Pt as a prototype *Philos. Mag. Lett.* **90** 851–9
[47] Du J, Shang S-L, Wang Y, Zhang A, Xiong S, Liu F and Liu Z-K 2021 Underpinned exploration for magnetic structure, lattice dynamics, electronic properties, and disproportionation of yttrium nickelate *AIP Adv.* **11** 015028
[48] Du J, Malyi O I, Shang S-L, Wang Y, Zhao X-G, Liu F, Zunger A and Liu Z-K 2022 Density functional thermodynamic description of spin, phonon and displacement degrees of freedom in antiferromagnetic-to-paramagnetic phase transition in YNiO3 *Mater. Today Phys.* **27** 100805
[49] Liu Z K, Wang Y and Shang S-L 2022 Zentropy Theory for Positive and Negative Thermal Expansion *J. Phase Equilibria Diffus.* **43** 598–605
[50] Evteev A V, Levchenko E V, Belova I V, Kozubski R, Liu Z K and Murch G E 2014 Thermotransport in binary system: case study on Ni 50 Al 50 melt *Philos. Mag.* **94** 3574–602
[51] Levchenko E V, Evteev A V, Ahmed T, Kromik A, Kozubski R, Belova I V, Liu Z-K and Murch G E 2016 Influence of the interatomic potential on thermotransport in binary liquid alloys: case study on NiAl *Philos. Mag.* **96**
[52] Ahmed T, Wang W Y, Kozubski R, Liu Z-K, Belova I V. and Murch G E 2018 Interdiffusion and thermotransport in Ni–Al liquid alloys *Philos. Mag.* **98** 2221–46
[53] Tang J, Xue X, Yi Wang W, Lin D, Ahmed T, Wang J, Tang B, Shang S, Belova I V., Song H, Murch G E, Li J and Liu Z K 2020 Activation volume dominated diffusivity of Ni50Al50 melt under extreme conditions *Comput. Mater. Sci.* **171** 109263
[54] Belova I V., Liu Z-K and Murch G E 2021 Exact phenomenological theory for thermotransport in a solid binary alloy *Philos. Mag. Lett.* **101** 123–31
[55] Gibbs J W 1873 Method of geometrical representation of the thermodynamic properties of substances by means of surfaces. *Trans. Connect. Acad. II* **December** 382–404
[56] Gibbs J W 1878 Abstract of "the equilibrium of heterogeneous substances" *Am. J. Sci.* **s3-16**, **Dec** 441–58
[57] Liu Z K and Wang Y 2016 *Computational Thermodynamics of Materials* (Cambridge: Cambridge University Press)
[58] Born M and Oppenheimer R 1927 Quantum theory of molecules *Ann. Phys.* **84** 457–84
[59] Ceperley D M and Alder B J 1980 Ground state of the electron gas by a stochastic method *Phys. Rev. Lett.* **45** 566–9
[60] Perdew J P and Zunger A 1981 Self-interaction correction to density-functional approximations for many-electron systems *Phys. Rev. B* **23** 5048–79
[61] Kresse G and Joubert D 1999 From ultrasoft pseudopotentials to the projector augmented-wave method *Phys. Rev. B* **59** 1758–75

[62] Shang S-L, Wang Y, Kim D and Liu Z-K 2010 First-principles thermodynamics from phonon and Debye model: Application to Ni and Ni3Al *Comput. Mater. Sci.* **47** 1040–8

[63] Wang Y, Liao M, Bocklund B J, Gao P, Shang S-L, Kim H, Beese A M, Chen L-Q and Liu Z-K 2021 DFTTK: Density Functional Theory ToolKit for high-throughput lattice dynamics calculations *CALPHAD* **75** 102355

[64] Anon DFTTK: Density Functional Theory Tool Kits *https://www.dfttk.org/*

[65] Wang Y, Shang S, Liu Z-K and Chen L-Q 2012 Mixed-space approach for calculation of vibration-induced dipole-dipole interactions *Phys. Rev. B* **85** 224303

[66] Wang Y, Chen L-Q and Liu Z-K 2014 YPHON: A package for calculating phonons of polar materials *Comput. Phys. Commun.* **185** 2950–68

[67] Wang Y, Zhang L A, Shang S L, Liu Z K and Chen L Q 2013 Accurate calculations of phonon dispersion in CaF2 and CeO2 *Phys. Rev. B* **88** 24304

[68] Liu Z K, Wang Y and Shang S 2014 Thermal Expansion Anomaly Regulated by Entropy *Sci. Rep.* **4** 7043

[69] Liu Z K 2024 On Gibbs Equilibrium and Hillert Nonequilibrium Thermodynamics *J. Phase Equilibria Diffus.* **Invited** ArXiv 2402.14231

[70] Coleman B D and Truesdell C 1960 On the reciprocal relations of Onsager *J. Chem. Phys.* **33** 28–31

[71] Truesdell C 1962 Mechanical basis of diffusion *J. Chem. Phys.* **37** 2336–44

[72] Andersson J and Ågren J 1992 Models for numerical treatment of multicomponent diffusion in simple phases *J. Appl. Phys.* **72** 1350–5

[73] de Groot S R and Mazur P 1984 *Non-equilibrium thermodynamics* (Dover Publications, Inc. New York)

[74] Kondepudi D and Prigogine I 2015 *Modern Thermodynamics: From Heat Engines to Dissipative Structures* (Hoboken, New Jersey: John Wiley & Sons Ltd.)

[75] Jou D, Casas-Vázquez J and Lebon G 2010 *Extended irreversible thermodynamics* (Springer Netherlands)

[76] Lebon G and Jou D 2015 Early history of extended irreversible thermodynamics (1953-1983): An exploration beyond local equilibrium and classical transport theory *Eur. Phys. J. H* **40** 205–40

[77] Agren J 1981 Computer simulations of diffusional reactions in multicomponent alloys with special applications to steel PhD Thesis

[78] Bothe D 2011 On the Maxwell-Stefan Approach to Multicomponent Diffusion *Parabolic Problems: The Herbert Amann Festschrift* ed J Escher, P Guidotti, M Hieber, P Mucha, J W Prüss, Y Shibata, G Simonett, C Walker and W Zajaczkowski (Basel: Springer Basel) pp 81–93

[79] Allie-Ebrahim T, Zhu Q, Bräuer P, Moggridge G D and D'Agostino C 2017 Maxwell–Stefan diffusion coefficient estimation for ternary systems: an ideal ternary alcohol system *Phys. Chem. Chem. Phys.* **19** 16071–7

[80] Anon Thermo-Calc Databases

[81] Anon CompuTherm Software and Databases

[82] Kirkaldy J S and Young D J 1987 *Diffusion in the condensed state* (Institute of Metals)

[83] Liu Z K, Höglund L, Jönsson B and Ågren J 1991 An experimental and theoretical study of cementite dissolution in an Fe-Cr-C alloy *Metall. Trans. A* **22** 1745–52

[84] Helander T and Ågren J 1997 Computer simulation of multicomponent diffusion in joints

of dissimilar steels *Metall. Mater. Trans. A* **28** 303–8
[85] Darken L S 1942 Diffusion in metal accompanied by phase change *Trans. Am. Inst. Min. Metall. Eng.* **150** 157–69
[86] Darken L S 1948 Diffusion, mobility and their interrelation through free energy in binary metallic systems *Trans. Am. Inst. Min. Metall. Eng.* **175** 184–201
[87] Darken L S 1949 Diffusion of carbon in austenite with a discontinuity in composition *Trans. Am. Inst. Min. Metall. Eng.* **180** 430–8
[88] Andersson J-O, Helander T, Höglund L, Shi P and Sundman B 2002 Thermo-Calc & DICTRA, computational tools for materials science *CALPHAD* **26** 273–312
[89] Hillert M 1961 A solid-solution model for inhomogeneous systems *Acta Metall.* **9** 525–35
[90] Cahn J W 1962 On spinodal decomposition in cubic crystals *Acta Metall.* **10** 179–83
[91] Chen L-Q 2002 Phase-field models for microstructure evolution *Annu. Rev. Mater. Res.* **32** 113–40
[92] Krishna R 2015 Uphill diffusion in multicomponent mixtures *Chem. Soc. Rev.* **44** 2812–36
[93] Kirkendall E, Thomassen L and Upthegrove C 1939 Rates of diffusion of copper and zinc in alpha brass. *Trans. Am. Inst. Min. Metall. Eng.* **133** 186–203
[94] Kirkendall E O 1942 Diffusion of zinc in alpha brass *Trans. Am. Inst. Min. Metall. Eng.* **147** 104–9
[95] Smigelskas A D and Kirkendall E O 1947 Zinc diffusion in alpha-brass *Trans. Am. Inst. Min. Metall. Eng.* **171** 130–42
[96] Höglund L and Ågren J 2001 Analysis of the Kirkendall effect, marker migration and pore formation *Acta Mater.* **49** 1311–7
[97] Campbell C E, Zhao J-C and Henry M F 2004 Comparison of Experimental and Simulated Multicomponent Ni-Base Superalloy Diffusion Couples *J. Phase Equilibria Diffus.* **25** 6–15
[98] Mantina M, Wang Y, Arroyave R, Chen L Q, Liu Z K and Wolverton C 2008 First-Principles Calculation of Self-Diffusion Coefficients *Phys. Rev. Lett.* **100** 215901
[99] Mantina M, Shang S L, Wang Y, Chen L Q and Liu Z K 2009 3 d transition metal impurities in aluminum: A first-principles study *Phys. Rev. B* **80** 184111
[100] Mantina M, Wang Y, Chen L Q, Liu Z K and Wolverton C 2009 First principles impurity diffusion coefficients *Acta Mater.* **57** 4102–8
[101] Hargather C Z, Shang S-L, Liu Z-K and Du Y 2014 A first-principles study of self-diffusion coefficients of fcc Ni *Comput. Mater. Sci.* **86** 17–23
[102] Hargather C Z, Shang S L and Liu Z K 2018 A comprehensive first-principles study of solute elements in dilute Ni alloys: Diffusion coefficients and their implications to tailor creep rate *Acta Mater.* **157** 126–41
[103] Mantina M, Chen L Q and Liu Z K 2009 Predicting Diffusion Coefficients from First-principles via Eyring's Reaction Rate Theory *Defect Diffus. Forum* **294** 1–13
[104] Ganeshan S, Shang S L, Zhang H, Wang Y, Mantina M and Liu Z K 2009 Elastic constants of binary Mg compounds from first-principles calculations *Intermetallics* **17** 313–8
[105] Zhou B-C, Shang S-L, Wang Y and Liu Z K 2016 Diffusion coefficients of alloying elements in dilute Mg alloys: A comprehensive first-principles study *Acta Mater.* **103** 573–86

[106] Shang S L, Hector L G, Wang Y and Liu Z K 2011 Anomalous energy pathway of vacancy migration and self-diffusion in hcp Ti *Phys. Rev. B* **83** 224104
[107] Eyring H 1935 The Activated Complex in Chemical Reactions *J. Chem. Phys.* **3** 107–15
[108] Vineyard G H 1957 Frequency factors and isotope effects in solid state rate processes *J. Phys. Chem. Solids* **3** 121–7
[109] Henkelman G, Uberuaga B P and Jónsson H 2000 A climbing image nudged elastic band method for finding saddle points and minimum energy paths *J. Chem. Phys.* **113** 9901–4
[110] Le Claire A D 1978 Solute diffusion in dilute alloys *J. Nucl. Mater.* **69–70** 70–96
[111] Wimmer E, Wolf W, Sticht J, Saxe P, Geller C B, Najafabadi R and Young G A 2008 Temperature-dependent diffusion coefficients from ab initio computations: Hydrogen, deuterium, and tritium in nickel *Phys. Rev. B* **77** 134305
[112] Höglund L and Ågren J 2010 Simulation of Carbon Diffusion in Steel Driven by a Temperature Gradient *J. Phase Equilibria Diffus.* **31** 212–5
[113] Szilard L 1929 Uber die Entropieverminderung in einem thermodynamischen System bei Eingriffen intelligenter Wesen *Zeitschrift fur Phys.* **53** 840–56
[114] Szilard L 1964 On the Decrease of Entropy in a Thermodynamic System by the Intervention of Intelligent Beings *Behav. Sci.* **9** 301–10
[115] Shannon C E 1948 A Mathematical Theory of Communication: Part III *Bell Syst. Tech. J.* **27** 623–56
[116] Shannon C E 1951 Prediction and Entropy of Printed English *Bell Syst. Tech. J.* **30** 50–64
[117] Zivieri R 2022 From Thermodynamics to Information: Landauer's Limit and Negentropy Principle Applied to Magnetic Skyrmions *Front. Phys.* **10** 769904
[118] van de Walle A, Tiwary P, de Jong M, Olmsted D L, Asta M, Dick A, Shin D, Wang Y, Chen L-Q and Liu Z-K 2013 Efficient stochastic generation of special quasirandom structures *CALPHAD* **42** 13–8
[119] Jiang C and Uberuaga B P 2016 Efficient Ab initio Modeling of Random Multicomponent Alloys *Phys. Rev. Lett.* **116** 105501
[120] Singh R, Sharma A, Singh P, Balasubramanian G and Johnson D D 2021 Accelerating computational modeling and design of high-entropy alloys *Nat. Comput. Sci.* **1** 54–61
[121] Hong Q-J and Liu Z K 2024 A generalized approach for rapid entropy calculation of liquids and solids ArXiv 2403.19872
[122] Krajewski A M, Siegel J W, Xu J and Liu Z-K 2022 Extensible Structure-Informed Prediction of Formation Energy with improved accuracy and usability employing neural networks *Comput. Mater. Sci.* **208** 111254
[123] Li K, DeCost B, Choudhary K, Greenwood M and Hattrick-Simpers J 2023 A critical examination of robustness and generalizability of machine learning prediction of materials properties *npj Comput. Mater.* **9** 55
[124] Fultz B 2010 Vibrational thermodynamics of materials *Prog. Mater. Sci.* **55** 247–352
[125] Guillaume C E 1897 Recherches sur les aciers au nickel. Dilatations aux temperatures elevees; resistance electrique *C. R. Acad. Sci. Paris* **125** 235–8
[126] GUILLAUME C-E 1933 Invar *Nature* **131** 658–658
[127] Wittenauer J 1997 *The Invar effect: A centennial symposium:* (The Minerals, Metals & Materials Society (TMS))
[128] Shang S, Zhang H, Ganeshan S and Liu Z-K 2008 The development and application of a thermodynamic database for magnesium alloys *JOM* **60** 45–7

[129] Zhang H, Wang Y, Shang S, Chen L-Q and Liu Z-K 2008 Thermodynamic modeling of Mg–Ca–Ce system by combining first-principles and CALPHAD method *J. Alloys Compd.* **463** 294–301
[130] Liu Z K, Wang Y and Shang S-L 2011 Origin of negative thermal expansion phenomenon in solids *Scr. Mater.* **65** 664–7
[131] Shang S-L, Wang Y and Liu Z K 2023 Quantifying the degree of disorder and associated phenomena in materials through zentropy: Illustrated with Invar Fe3Pt *Scr. Mater.* **225** 115164
[132] Shang S-L, Wang Y and Liu Z-K 2010 Thermodynamic fluctuations between magnetic states from first-principles phonon calculations: The case of bcc Fe *Phys. Rev. B* **82** 014425
[133] Shang S L, Saal J E, Mei Z G, Wang Y and Liu Z K 2010 Magnetic thermodynamics of fcc Ni from first-principles partition function approach *J. Appl. Phys.* **108** 123514
[134] Wang Y, Shang S L, Hui X D, Chen L Q and Liu Z K 2010 Effects of spin structures on phonons in BaFe2As2 *Appl. Phys. Lett.* **97** 022504
[135] Wang Y, Saal J E, Shang S L, Hui X D, Chen L Q and Liu Z K 2011 Effects of spin structures on Fermi surface topologies in BaFe2As2 *Solid State Commun.* **151** 272–5
[136] Wang Y, Shang S L, Chen L Q and Liu Z K 2011 Magnetic excitation and thermodynamics of BaFe2As2 *Int. J. Quantum Chem.* **111** 3565–70
[137] Liu Z K, Shang S-L, Du J and Wang Y 2023 Parameter-free prediction of phase transition in PbTiO3 through combination of quantum mechanics and statistical mechanics *Scr. Mater.* **232** 115480
[138] Hew N L E, Shang S-L and Liu Z-K 2024 Predicting Phase Transitions in PbTiO3 using Zentropy
[139] Krisch M, Farber D L, Xu R, Antonangeli D, Aracne C M, Beraud A, Chiang T C, Zarestky J, Kim D Y, Isaev E I, Ahuja R and Johansson B 2011 Phonons of the anomalous element cerium *Proc. Natl. Acad. Sci.* **108** 9342–5
[140] Liu Z K, Hew N L E and Shang S-L 2024 Zentropy theory for accurate prediction of free energy, volume, and thermal expansion without fitting parameters *Microstructures* **4** 2024009
[141] Liu Z K 2022 DE-SC0023185: Zentropy Theory for Transformative Functionalities of Magnetic and Superconducting Materials *DE-SC0023185*
[142] Liu Z K and Shang S-L 2024 Revealing Symmetry-broken Superconducting Configurations by Density Functional Theory ArXiv 2404.00719
[143] Lau B and Millis A J 2013 Theory of the Magnetic and Metal-Insulator Transitions in RNiO3 Bulk and Layered Structures *Phys. Rev. Lett.* **110** 126404
[144] Middey S, Chakhalian J, Mahadevan P, Freeland J W, Millis A J and Sarma D D 2016 Physics of Ultrathin Films and Heterostructures of Rare-Earth Nickelates *Annu. Rev. Mater. Res.* **46** 305–34
[145] Zunger A, Wei S H, Ferreira L G and Bernard J E 1990 Special Quasirandom Structures *Phys. Rev. Lett.* **65** 353–6
[146] Jiang C, Wolverton C, Sofo J, Chen L Q and Liu Z K 2004 First-principles study of binary bcc alloys using special quasirandom structures *Phys. Rev. B* **69** 214202
[147] Evans D J, Cohen E G D and Morriss G P 1993 Probability of second law violations in shearing steady states *Phys. Rev. Lett.* **71** 2401–4

[148] Jarzynski C 1997 Nonequilibrium Equality for Free Energy Differences *Phys. Rev. Lett.* **78** 2690–3

[149] Wang G M, Sevick E M, Mittag E, Searles D J and Evans D J 2002 Experimental Demonstration of Violations of the Second Law of Thermodynamics for Small Systems and Short Time Scales *Phys. Rev. Lett.* **89** 050601

[150] Jarzynski C 2011 Equalities and Inequalities: Irreversibility and the Second Law of Thermodynamics at the Nanoscale *Annu. Rev. Condens. Matter Phys.* **2** 329–51

[151] Sagawa T and Ueda M 2012 Fluctuation Theorem with Information Exchange: Role of Correlations in Stochastic Thermodynamics *Phys. Rev. Lett.* **109** 180602

[152] Maillet O, Erdman P A, Cavina V, Bhandari B, Mannila E T, Peltonen J T, Mari A, Taddei F, Jarzynski C, Giovannetti V and Pekola J P 2019 Optimal Probabilistic Work Extraction beyond the Free Energy Difference with a Single-Electron Device *Phys. Rev. Lett.* **122** 150604

[153] Seifert U 2020 Entropy and the second law for driven, or quenched, thermally isolated systems *Phys. A* **552** 121822

[154] Anon Bird and Cage Thaumatrope

[155] Sicron N, Ravel B, Yacoby Y, Stern E A, Dogan F and Rehr J J 1994 Nature of the ferroelectric phase transition in PbTiO3 *Phys. Rev. B* **50** 13168–80

[156] Sicron N, Ravel B, Yacoby Y, Stern E A, Dogan F and Rehr J J 1995 The ferroelectric phase transition in PbTiO3 from a local perspective *Phys. B* **208–209** 319–20

[157] Ravel B, Sicron N, Yacoby Y, Stern E A, Dogan F, Rehr J J, Slcron N, Yacoby Y, Stern E A, Dogan F and Rehr J J 1995 Order-disorder behavior in the phase transition of PbTiO3 *Ferroelectrics* **164** 265–77

[158] Fang H, Wang Y, Shang S and Liu Z K 2015 Nature of ferroelectric-paraelectric phase transition and origin of negative thermal expansion in PbTiO3 *Phys. Rev. B* **91** 024104

[159] Shirane G and Hoshino S 1951 On the phase transition in lead titanate *J. Phys. Soc. Japan* **6** 265–70

[160] Wang Y, Hu Y-J, Bocklund B, Shang S-L, Zhou B-C, Liu Z K and Chen L-Q 2018 First-principles thermodynamic theory of Seebeck coefficients *Phys. Rev. B* **98** 224101

[161] Wang Y, Chong X, Hu Y J, Shang S L, Drymiotis F R, Firdosy S A, Star K E, Fleurial J P, Ravi V A, Chen L Q and Liu Z K 2019 An alternative approach to predict Seebeck coefficients: Application to La 3−x Te 4 *Scr. Mater.* **169** 87–91

[162] Iacopini S and Piazza R 2003 Thermophoresis in protein solutions *Europhys. Lett.* **63** 247–53

[163] Kita R, Polyakov P and Wiegand S 2007 Ludwig−Soret Effect of Poly( N -isopropylacrylamide): Temperature Dependence Study in Monohydric Alcohols *Macromolecules* **40** 1638–42

[164] Kishikawa Y, Wiegand S and Kita R 2010 Temperature Dependence of Soret Coefficient in Aqueous and Nonaqueous Solutions of Pullulan *Biomacromolecules* **11** 740–7

[165] Iacopini S, Rusconi R and Piazza R 2006 The “macromolecular tourist": Universal temperature dependence of thermal diffusion in aqueous colloidal suspensions *Eur. Phys. J. E* **19** 59–67

[166] de Gans B-J, Kita R, Wiegand S and Luettmer-Strathmann J 2003 Unusual Thermal Diffusion in Polymer Solutions *Phys. Rev. Lett.* **91** 245501

[167] Costesèque P and Loubet J-C 2003 Measuring the Soret coefficient of binary

hydrocarbon mixtures in packed thermogravitational columns (contribution of Toulouse University to the benchmark test) *Philos. Mag.* **83** 2017–22

[168] Hartmann S, Wittko G, Köhler W, Morozov K I, Albers K and Sadowski G 2012 Thermophobicity of Liquids: Heats of Transport in Mixtures as Pure Component Properties *Phys. Rev. Lett.* **109** 065901

[169] Schraml M, Bataller H, Bauer C, Bou-Ali M M, Croccolo F, Lapeira E, Mialdun A, Möckel P, Ndjaka A T, Shevtsova V and Köhler W 2021 The Soret coefficients of the ternary system water/ethanol/triethylene glycol and its corresponding binary mixtures *Eur. Phys. J. E* **44** 128

[170] Rahman M A and Saghir M Z 2014 Thermodiffusion or Soret effect: Historical review *Int. J. Heat Mass Transf.* **73** 693–705

[171] Tu K N 2003 Recent advances on electromigration in very-large-scale-integration of interconnects *J. Appl. Phys.* **94** 5451

[172] Chen C, Tong H M and Tu K N 2010 Electromigration and Thermomigration in Pb-Free Flip-Chip Solder Joints *Annu. Rev. Mater. Res.* **40** 531–55

[173] Tu K N, Liu Y and Li M 2017 Effect of Joule heating and current crowding on electromigration in mobile technology *Appl. Phys. Rev.* **4** 011101

[174] Tu K N and Gusak A N 2020 Mean-Time-To-Failure Equations for Electromigration, Thermomigration, and Stress Migration *IEEE Trans. Components, Packag. Manuf. Technol.* **10** 1427–31

[175] Kirchheim R 1992 Stress and electromigration in Al-lines of integrated circuits *Acta Metall. Mater.* **40** 309–23

[176] Basaran C, Lin M and Ye H 2003 A thermodynamic model for electrical current induced damage *Int. J. Solids Struct.* **40** 7315–27

[177] Wiseman G G and Kuebler J K 1963 Electrocaloric Effect in Ferroelectric Rochelle Salt *Phys. Rev.* **131** 2023–7

[178] Lombardo G and Pohl R O 1965 Electrocaloric Effect and a New Type of Impurity Mode *Phys. Rev. Lett.* **15** 291–3

[179] Lu S-G and Zhang Q 2009 Electrocaloric Materials for Solid-State Refrigeration *Adv. Mater.* **21** 1983–7

[180] Scott J F 2011 Electrocaloric Materials *Annu. Rev. Mater. Res.* **41** 229–40

[181] Moya X, Kar-Narayan S and Mathur N D 2014 Caloric materials near ferroic phase transitions *Nat. Mater.* **13** 439–50

[182] Moya X and Mathur N D 2020 Caloric materials for cooling and heating *Science* **370** 797–803

[183] Qian X, Han D, Zheng L, Chen J, Tyagi M, Li Q, Du F, Zheng S, Huang X, Zhang S, Shi J, Huang H, Shi X, Chen J, Qin H, Bernholc J, Chen X, Chen L-Q, Hong L and Zhang Q M 2021 High-entropy polymer produces a giant electrocaloric effect at low fields *Nature* **600** 664–9

[184] Caspari M E and Merz W J 1950 The Electromechanical Behavior of BaTiO3 Single-Domain Crystals *Phys. Rev.* **80** 1082–9

[185] Kulcsar F 1959 Electromechanical Properties of Lead Titanate Zirconate Ceramics Modified with Certain Three-or Five-Valent Additions *J. Am. Ceram. Soc.* **42** 343–9

[186] Somlyo A V and Somlyo A P 1968 Electromechanical and pharmacomechanical coupling in vascular smooth muscle. *J. Pharmacol. Exp. Ther.* **159** 129–45

[187] Zhao J, Zhang Q M, Kim N and Shrout T 1995 Electromechanical Properties of Relaxor Ferroelectric Lead Magnesium Niobate-Lead Titanate Ceramics *Jpn. J. Appl. Phys.* **34** 5658–63

[188] Park S-E and Shrout T R 1997 Ultrahigh strain and piezoelectric behavior in relaxor based ferroelectric single crystals *J. Appl. Phys.* **82** 1804–11

[189] Fu H and Cohen R E 2000 Polarization rotation mechanism for ultrahigh electromechanical response in single-crystal piezoelectrics *Nature* **403** 281–3

[190] Kutnjak Z, Petzelt J and Blinc R 2006 The giant electromechanical response in ferroelectric relaxors as a critical phenomenon *Nature* **441** 956–9

[191] Ahart M, Somayazulu M, Cohen R E, Ganesh P, Dera P, Mao H, Hemley R J, Ren Y, Liermann P and Wu Z 2008 Origin of morphotropic phase boundaries in ferroelectrics *Nature* **451** 545–8

[192] Li F, Cabral M J, Xu B, Cheng Z, Dickey E C, LeBeau J M, Wang J, Luo J, Taylor S, Hackenberger W, Bellaiche L, Xu Z, Chen L-Q, Shrout T R and Zhang S 2019 Giant piezoelectricity of Sm-doped Pb(Mg 1/3 Nb 2/3 )O 3 -PbTiO 3 single crystals *Science* **364** 264–8

*Table 1: Physical quantities related to 1st directives between molar quantities (first column) and potentials (first row), slightly modified from Ref.* [33].

| | *T*, Temperature | $\boldsymbol{\sigma}$, Stress | $\boldsymbol{E}$, Electrical field | $\boldsymbol{\mathcal{H}}$, Magnetic field | $\mu_i$, Chemical potential |
|---|---|---|---|---|---|
| *S*, Entropy | Heat capacity | Piezocaloric effect | Electrocaloric effect | Magnetocaloric effect | $\frac{\partial S}{\partial \mu_k}$ |
| $\boldsymbol{\varepsilon}$, Strain | Thermal expansion | Elastic compliance | Converse piezoelectricity | Piezomagnetic moduli | $\frac{\partial \varepsilon_{ij}}{\partial \mu_k}$ |
| $\boldsymbol{\theta}$, Electric displacement | Pyroelectric coefficients | Piezoelectric moduli | Permittivity | Magnetoelectric coefficient | $\frac{\partial D_i}{\partial \mu_k}$ |
| $\boldsymbol{B}$, Magnetic induction | Pyromagnetic coefficient | Piezomagnetic moduli | Electromagnetic coefficient | Permeability | $\frac{\partial B_i}{\partial \mu_k}$ |
| $N_j$, Moles | *Thermoreactivity* | *Stressoreactivity* | *Electroreactivity* | *Magnetoreactivity* | $\frac{\partial N_i}{\partial \mu_k}$, *Thermodynamic factor* |

*Table 2: Cross phenomenon coefficients represented by 1st derivatives between potentials, slightly modified from Ref. [33].*

| | $T$, Temperature | $\boldsymbol{\sigma}$, Stress | $\boldsymbol{E}$, Electrical field | $\boldsymbol{\mathcal{H}}$, Magnetic field | $\mu_i$, Chemical potential |
|---|---|---|---|---|---|
| $T$ | 1 | $-\frac{\partial S}{\partial \boldsymbol{\varepsilon}}$ | $-\frac{\partial S}{\partial \boldsymbol{\theta}}$ | $-\frac{\partial S}{\partial \boldsymbol{B}}$ | $-\frac{\partial S}{\partial c_i}$ Partial entropy |
| $\boldsymbol{\sigma}$ | $\frac{\partial \boldsymbol{\sigma}}{\partial T}$, Thermostress | 1 | $-\frac{\partial \boldsymbol{\varepsilon}}{\partial \boldsymbol{\theta}}$ | $-\frac{\partial \boldsymbol{\varepsilon}}{\partial \boldsymbol{B}}$ | $-\frac{\partial \boldsymbol{\varepsilon}}{\partial c_i}$ Partial strain |
| $\boldsymbol{E}$ | $\frac{\partial \boldsymbol{E}}{\partial T}$, Thermoelectric | $\frac{\partial \boldsymbol{E}}{\partial \sigma}$, Piezoelectric | 1 | $-\frac{\partial \boldsymbol{\theta}}{\partial \boldsymbol{B}}$ | $-\frac{\partial \boldsymbol{\theta}}{\partial c_i}$ Partial electrical displacement |
| $\boldsymbol{\mathcal{H}}$ | $\frac{\partial \boldsymbol{\mathcal{H}}}{\partial T}$, Thermomagnetic | $\frac{\partial \boldsymbol{\mathcal{H}}}{\partial \sigma}$, Piezomagnetic | $\frac{\partial \boldsymbol{\mathcal{H}}}{\partial \boldsymbol{E}}$, Electromagnetic | 1 | $-\frac{\partial B}{\partial c_i}$ Partial magnetic induction |
| $\mu_i$ | $\frac{\partial \mu_i}{\partial T}$ Thermodiffusion | $\frac{\partial \mu_i}{\partial \boldsymbol{\sigma}}$ Stressmigration | $\frac{\partial \mu_i}{\partial \boldsymbol{E}}$ Electromigration | $\frac{\partial \mu_i}{\partial \boldsymbol{\mathcal{H}}}$ Magnetomigration | $\frac{\partial \mu_i}{\partial \mu_j} = -\frac{\partial c_j}{\partial c_i} = \frac{\Phi_{ii}}{\Phi_{ji}}$ Crossdiffusion |

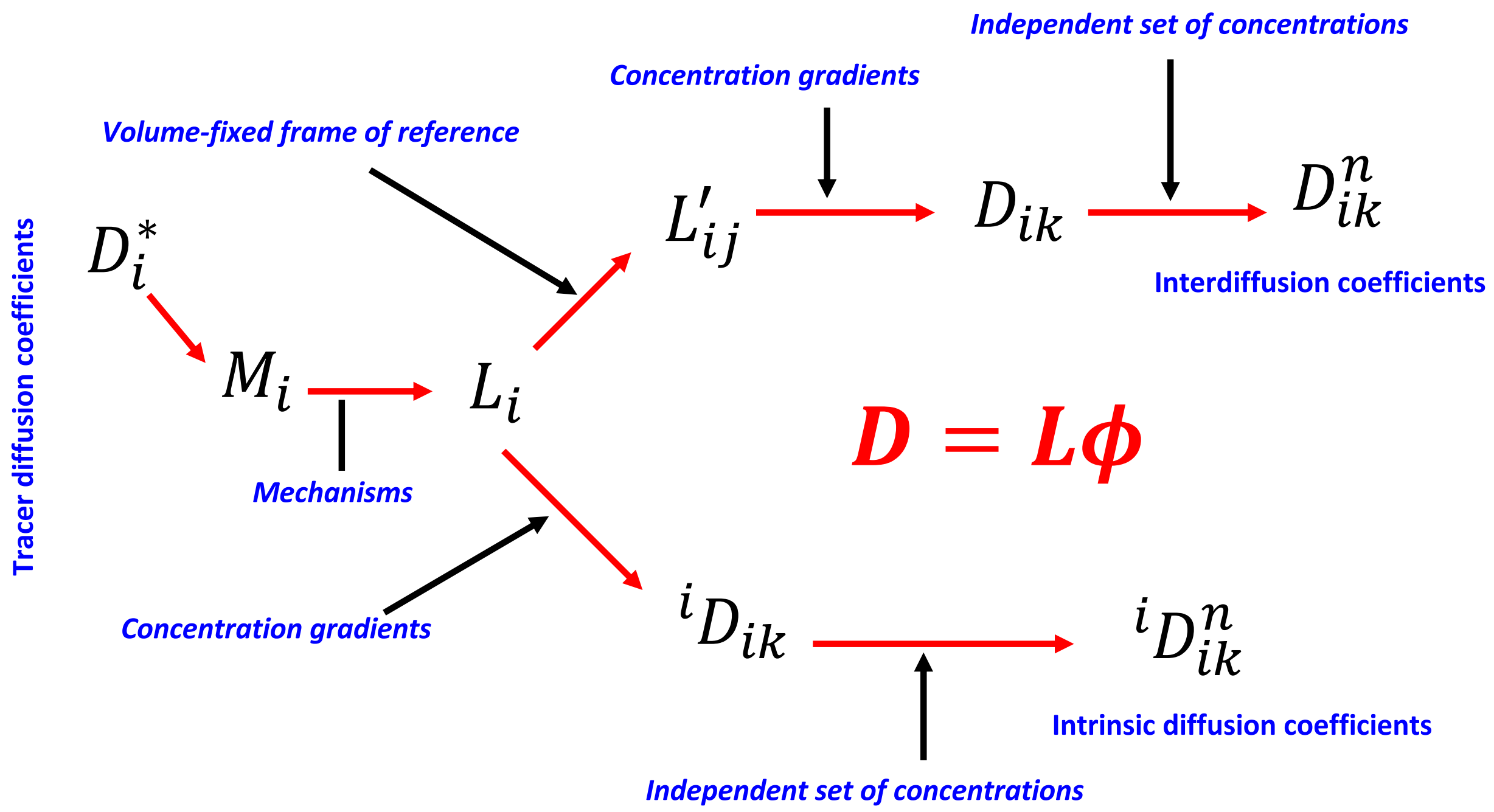

Figure 1

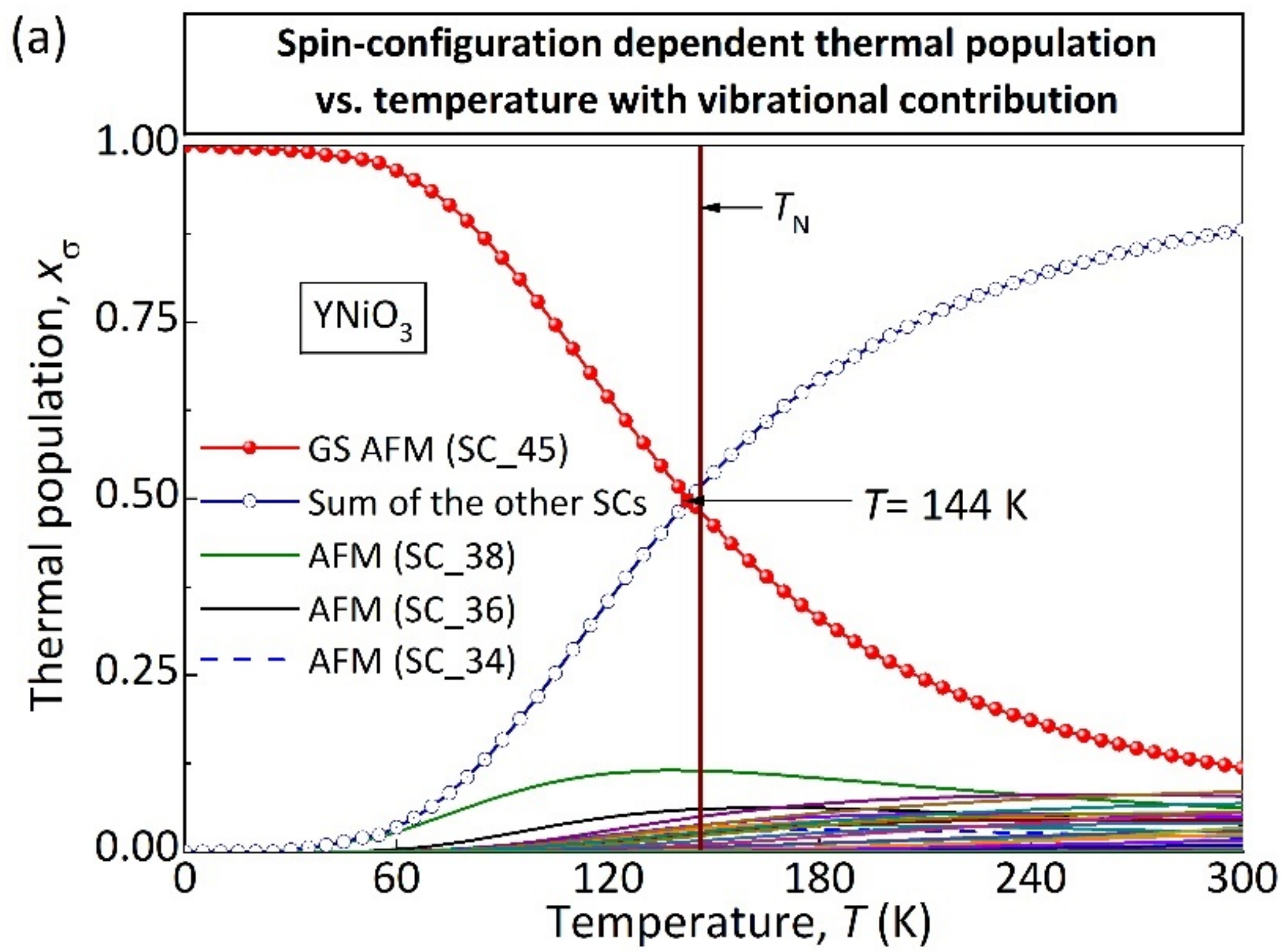

(i)

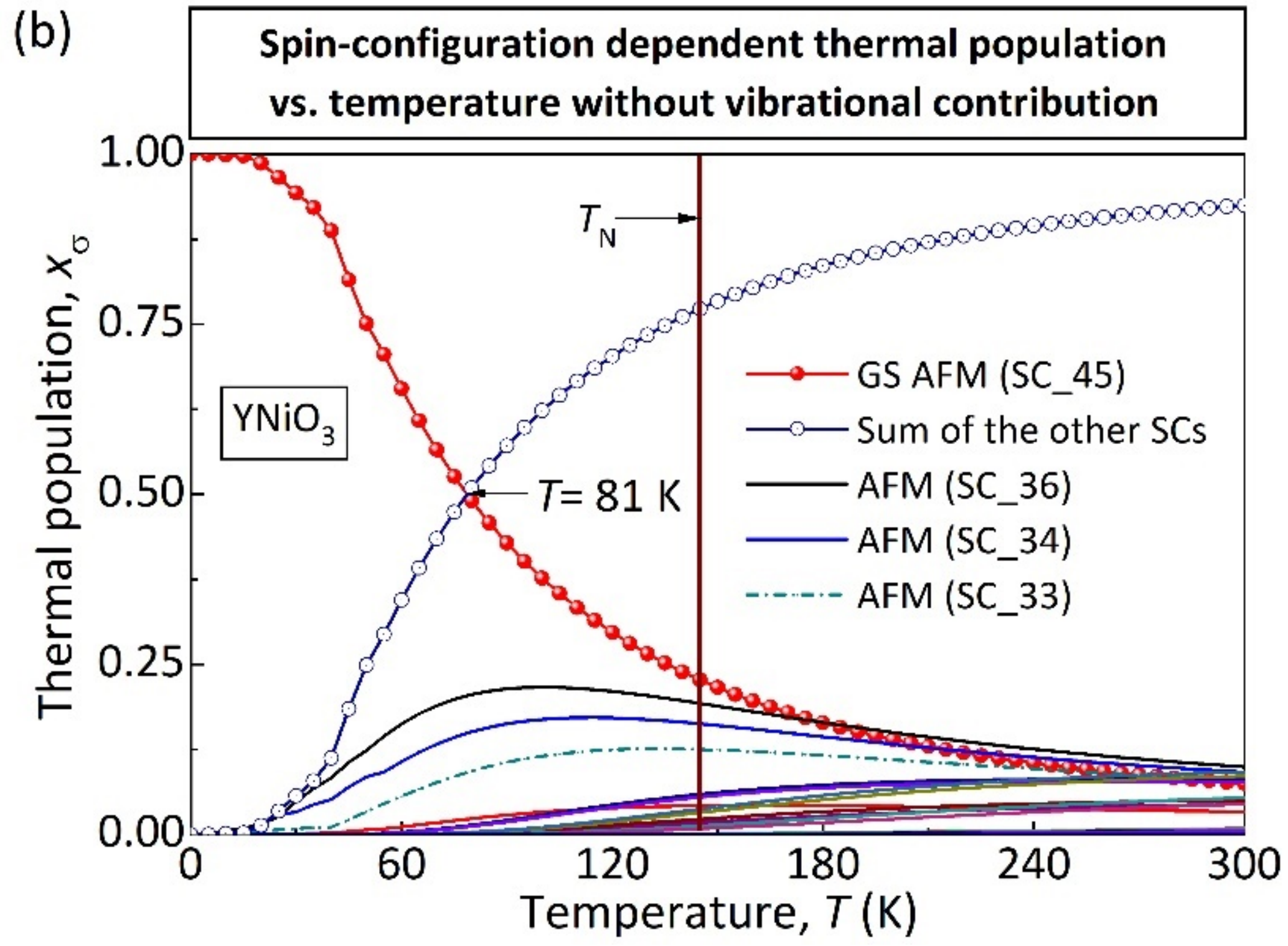

(ii)

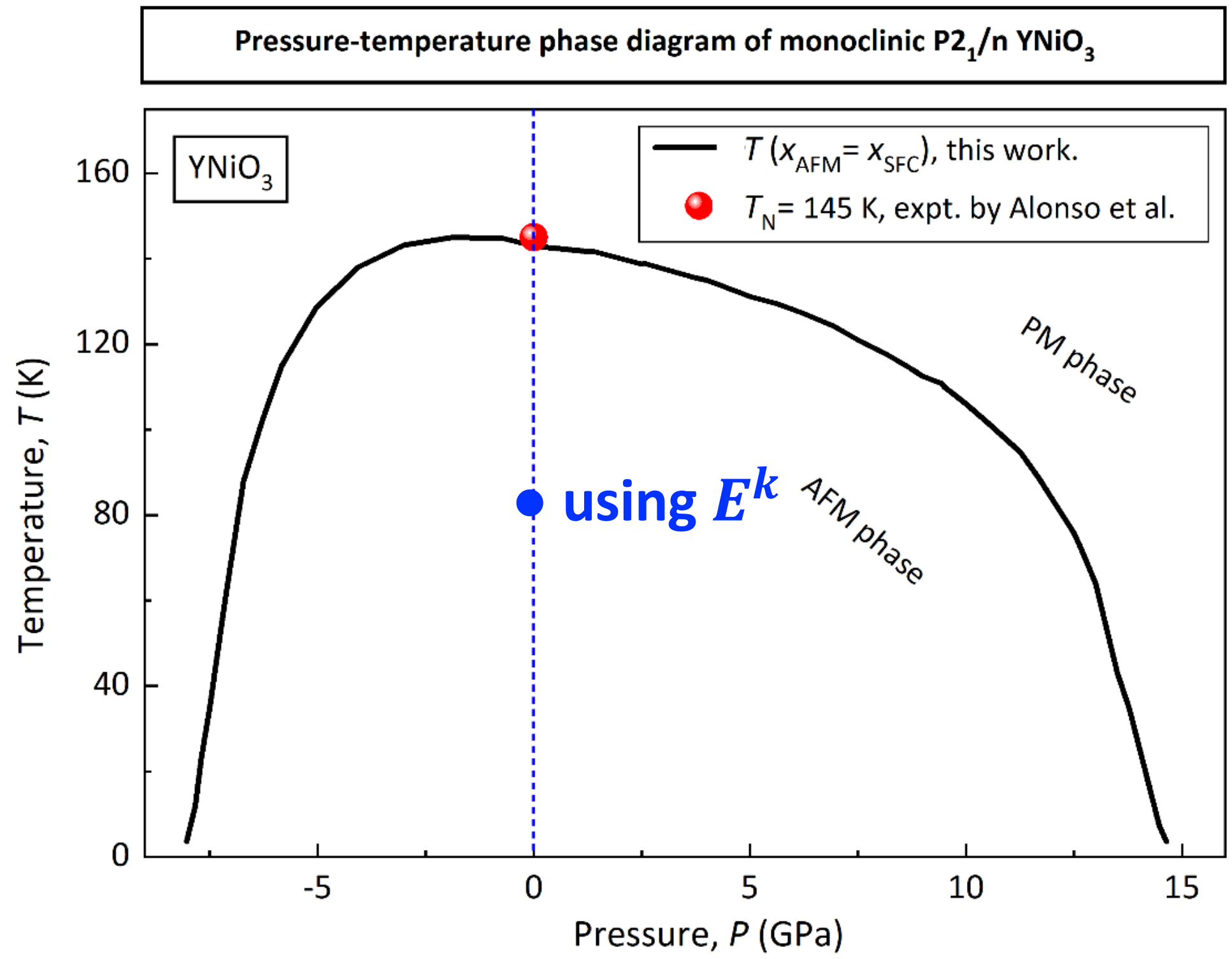

(iii)

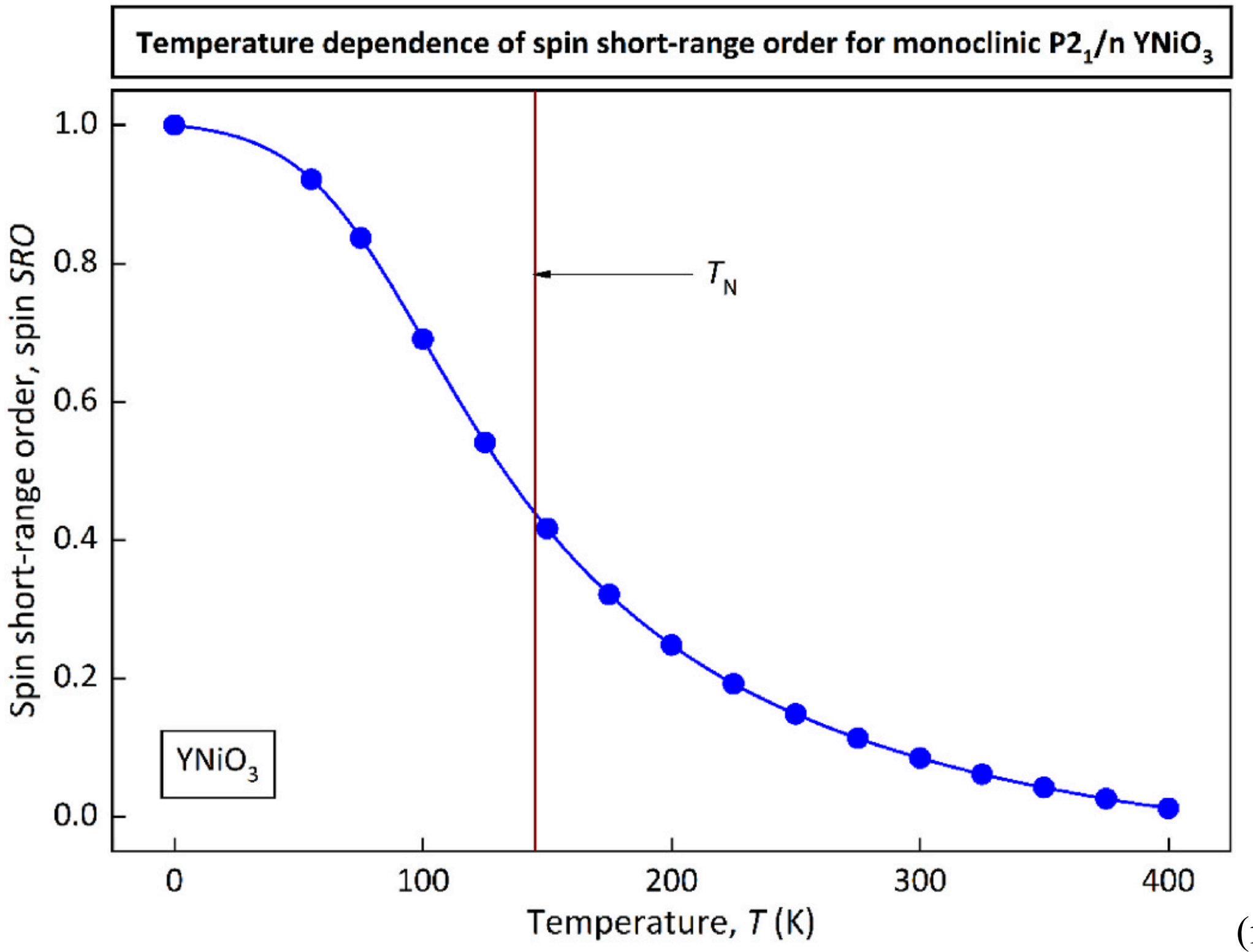

(iv)

Figure 2